\documentclass{article}

\usepackage{arxiv}

\usepackage[utf8]{inputenc} 
\usepackage[T1]{fontenc}    
\usepackage{lmodern}        
\usepackage{hyperref}       
\usepackage{url}            
\usepackage{booktabs}       
\usepackage{amsfonts}       
\usepackage{nicefrac}       
\usepackage{microtype}      
\usepackage{graphicx}

\title{Fixed and adaptive landmark sets\linebreak for finite
pseudometric spaces}

\author{
    Jason Cory Brunson
    \thanks{Supported by the University of Connecticut Skeletal,
Craniofacial and Oral Biology Training Grant (National Institute of
Dental and Craniofacial Research grant 5T90DE021989-07).}
   \\
    Laboratory for Systems Medicine \\
    University of Florida \\
  Gainesville, FL, USA \\
  \texttt{\href{mailto:jason.brunson@medicine.ufl.edu}{\nolinkurl{jason.brunson@medicine.ufl.edu}}} \\
   \And
    Yara Skaf
    \thanks{Supported by the University of Florida and Florida State
University Clinical and Translational Science Awards (National Center
for Advancing Translational Sciences grants TL1TR001428 and
UL1TR001427).}
   \\
    Laboratory for Systems Medicine \\
    University of Florida \\
  Gainesville, FL, USA \\
  \texttt{\href{mailto:yara.skaf@ufl.edu}{\nolinkurl{yara.skaf@ufl.edu}}} \\
  }

\providecommand{\tightlist}{%
  \setlength{\itemsep}{0pt}\setlength{\parskip}{0pt}}

\newlength{\cslhangindent}
\newlength{\csllabelwidth}
\newlength{\cslentryspacingunit} 
  {}%
  {\par}
\newenvironment{CSLReferences}[2] 
 {
  \setlength{\parindent}{0pt}
  \ifodd #1
  \let\oldpar\par
  \def\par{\hangindent=\cslhangindent\oldpar}
  \fi
  \setlength{\parskip}{#2\cslentryspacingunit}
 }%
 {}
\usepackage{calc}

\usepackage{format}
\begin{document}
\maketitle

\begin{abstract}
Topological data analysis (TDA) is an expanding field that leverages
principles and tools from algebraic topology to quantify structural
features of data sets or transform them into more manageable forms. As
its theoretical foundations have been developed, TDA has shown promise
in extracting useful information from high-dimensional, noisy, and
complex data such as those used in biomedicine. To improve efficiency,
these techniques may employ landmark samplers. The heuristic maxmin
procedure obtains a roughly even distribution of sample points by
implicitly constructing a cover comprising sets of uniform radius.
However, issues arise with data that vary in density or include points
with multiplicities, as are common in biomedicine. We propose an
analogous procedure, ``lastfirst'' based on ranked distances, which
implies a cover comprising sets of uniform cardinality. We first
rigorously define the procedure and prove that it obtains landmarks with
desired properties. We then perform benchmark tests and compare its
performance to that of maxmin, on feature detection and class prediction
tasks involving simulated and real-world biomedical data. Lastfirst is
more general than maxmin in that it can be applied to any data on which
arbitrary (and not necessarily symmetric) pairwise distances can be
computed. Lastfirst is more computationally costly, but our
implementation scales at the same rate as maxmin. We find that lastfirst
achieves comparable performance on prediction tasks and outperforms
maxmin on homology detection tasks. Where the numerical values of
similarity measures are not meaningful, as in many biomedical contexts,
lastfirst sampling may also improve interpretability.
\end{abstract}


\hypertarget{acknowledgments}{%
\section*{Acknowledgments}\label{acknowledgments}}
\addcontentsline{toc}{section}{Acknowledgments}

We are grateful to Reinhard Laubenbacher, Matthew Wheeler, and Peter
Bubenik for critical feedback on earlier drafts of this manuscript.

\pagebreak

\hypertarget{introduction}{%
\section{Introduction}\label{introduction}}

Topological data analysis (TDA) is a maturing field in data science at
the interface of statistics, computer science, and mathematics. Topology
is the discipline at the intersection of geometry (the study of shape)
and analysis (the study of continuity) that focuses on geometric
properties that are preserved under continuous transformations. TDA
consists of the use of computational theories of continuity to
investigate or exploit the structure of data. While TDA is most commonly
associated with persistent homology (PH), mapper-like constructions, and
non-linear dimension reduction (NLDR), it can be understood to include
such classical and conventional techniques as cluster analysis, network
analysis, and nearest neighbors prediction.

TDA methods have been deployed widely in biomedicine; see Skaf and
Laubenbacher (2022) for a review of applications of PH and mapper and
various reviews of applications of NLDR (Ivakhno and Armstrong 2007;
Reutlinger and Schneider 2012; Aziz et al. 2017; Viswanath et al. 2017;
Konstorum et al. 2018; Becht et al. 2019). These methods either require
that data be transformed into Euclidean vectors or perform such a
transformation, and for many types of biomedical data, including most
image and omics data, this is a straightforward requirement and yields
point clouds with desirable properties (separability). In contrast,
analysis tasks in clinical and public health often involve data that
have diverse numerical and categorical variable types, high rates and
complex patterns of missingness, or only a small number of variables
taking finitely many values. A variety of approaches have been taken to
apply classical topological methods to clinical and public health data,
whether by transformaing records to coordinate vectors (``vector space
embeddings'') or by using non-metric similarity measures such as those
common in ecology (Johnston 1976; Lee, Maslove, and Dubin 2015; Dai,
Zhu, and Liu 2020). Extensions of these ideas to the present-day TDA
toolkit are needed to bring more advanced methods to bear in domains
that rely on more challenging data.

One support tool for TDA is the selection of landmarks from a data set,
which can reduce the conceptual or computational complexity of an
analysis. Maxmin is an agglomerative procedure to sample landmarks that
are close to the data but well-separated from each other; it maintains a
constant lower bound on the ratio of the minimum distance between
landmarks to the maximum distance from the landmarks to the data
(Boissonnat, Chazal, and Yvinec 2018, \S\nbs{5.1.3}). It has been used
to expedite the calculation of PH (de Silva and Carlsson 2004) and of
mapper (Singh, Mémoli, and Carlsson 2007) and directly to compute
computational-topological representations of data. The procedure has the
advantage of being deterministic, being computationally efficient, and
generating more evenly distributed samples than random selection. In
addition to approximating PH, maxmin was used in these cases to reduce
the sizes of simplicial complex models of point cloud data for the sake
of visualization and exploration. Like other TDA tools, maxmin is most
useful on separable data and most efficient on data stored as Euclidean
coordinates.

In this paper, we describe an adaptation of maxmin to health data that
are not separable or Euclidean. Our procedure is deterministic and based
not on a raw similarity or distance measure but on a rank-order of
neighboring cases to each index case. We focus on three questions: (1)
How is maxmin most naturally modified to sample landmarks from health
data? (2) What useful properties can the modified procedure and its
samples and covers be ensured to have? (3) How does the procedure
perform on common analysis tasks on real-world data, and is its
performance comparable or superior to that of maxmin?

We organize the paper as follows: In the remainder of this section, we
introduce mathematical notation and use a simple example to motivate the
procedure. We define and prove algorithms and other properties in
Section\nbs\ref{sec:methods}, after which we describe our implementation
and several experiments. In Section\nbs\ref{sec:results} we summarize
key definitions, algorithms, and properties, report benchmark tests and
robustness checks, and report experimental results. We interpret our
findings and comment on limitations and ongoing work in
Section\nbs\ref{sec:discussion}.

\hypertarget{conventions}{%
\subsection{Conventions}\label{conventions}}

\((X, d_X)\) will refer to a finite pseudometric space with point set
\(X\) and pseudometric \(d_X:X\times X\to\bbR_{\geq 0}\), which by
definition satisfies all of the properties of a metric except that
\(d_X(x,y)=0\) implies \(x=y\). \((X,d_X)\) may be shortened to \(X\),
and \(d_X\) to \(d\), when clear from context. If \(x\neq y\) but
\(d(x,y)=0\) then \(x\) and \(y\) are said to be indistinguishable or
co-located. The cardinality of \(Y\subseteq X\) (counting
multiplicities) is denoted \(\abs{Y}\), the set of points co-located
with those in \(Y\), equivalent in this setting to the closure of \(Y\),
is denoted \(\cl{Y}\), and the set of equivalence classes of \(Y\) under
co-location is denoted \(\supp{Y}\). Throughout, let \(N=\abs{X}\). When
\(Y,Z\subseteq X\), let \(Y\wo Z\) denote the set difference
\(\{x \where x \in Y \wedge x \notin Z\}\). Then \(Y \wo \cl{Z}\) is the
set of points in \(Y\) (with multiplicities) that are distinguishable
from all points in \(Z\). This means that, when defined,
\(\min_{y\in Y\wo \cl{Z},z\in Z}{d(y, z)}>0\).

We denote the diameter \(D(Y)=\max_{x,y\in Y}{d(x,y)}\) and write:
\begin{align*}
d(Y,Z) &= \min_{y\in Y,z\in Z}{d(y,z)} & D(Y,Z) &= \max_{y\in Y,z\in Z}{d(y,z)} \\
d(x,Y) &= d(\{x\},Y)                   & D(x,Y) &= D(\{x\},Y)
\end{align*} If, for any \(x,y,z,w \in X\), \(d(x,y)=d(z,w)\) implies
\(\{x,y\}=\{z,w\}\)---that is, if no two pairs of points in \(X\) have
equal distance---then \(X\) is said to be in general position. We also
say that \(X\) is in \emph{locally general position} if, for any
\(x,y,z \in X\), \(d(x,y)=d(x,z)\) implies \(y=z\)---a weaker condition,
since there may exist \(w \in X\) for which \(d(x,y)=d(z,w)\) but
\(\{x,y\}\neq\{z,w\}\). Either condition implies that \(X\) is
Hausdorff: \(d(x,y)=0\) implies \(x=y\). \(f:X \to Y\) will denote a
morphism of pseudometric spaces, which we take to be a \(1\)-Lipschitz
map: \(d_X(x,y)\geq d_Y(f(x),f(y))\).

Denote by \(\power{X}\) the power set of \(X\) and by \(\order{X}\) the
set of ordered, non-duplicative sequences from \(X\). We use the ball
notation \(B_{\eps}(x) \in \power{X}\) for the set of points less than
distance \(\eps\) from a point \(x\); that is,
\(B_{\eps}(x) = \{ y \where d(x,y) < \eps \}\). We use an overline to
also include points exactly distance \(\eps\) from \(x\):
\(\cl{B_{\eps}}(x) = \{ y \where d(x,y) \leq \eps \}\). If \(k\) is an
integer such that \(\abs{\cl{B_\eps}(x)} \geq k\) while
\(\eps'<\eps \implies \abs{\cl{B_{\eps'}}(x)} < k\), then we call
\(N_k(x) = \cl{B_\eps}(x)\) the \(k\)-nearest neighborhood of \(x\).
When \(X\) is in locally general position, \(\abs{N_k(x)} = k\).

For convenience, we assume \(0\in\bbN\). For \(a,b \in \bbN\) with
\(a<b\), we use \([a,b]\) to denote the arithmetic sequence
\((a,a+1,\ldots,b)\). For \(a,b \in \bbN\), we use \(a^b\) to denote the
sequence \((a,\ldots,a)\) of length \(b\).

\hypertarget{motivation}{%
\subsection{Motivation}\label{motivation}}

In contrast to most geometric data analytic tools, such as linear
regression and principal components analysis, that reduce the
dimensionality of data, much TDA relies on the reduction of
\emph{cardinality}. As distinguished by Byczkowska-Lipińska and Wosiak
(2017), an \(n\times p\) data table of \(n\) cases (rows) and \(p\)
variables (columns) can be dimension-reduced to an \(n\times q\) table,
where \(q<p\), or cardinality-reduced to an \(m\times p\) table, where
\(m<n\). The most common cardinality reduction technique is data
reduction, and unsupervised clustering methods are classical examples.
Many popular TDA techniques reduce cardinality rather than dimension to
improve efficiency, often through landmark sampling: de Silva and
Carlsson (2004) propose witness complexes, related to alpha complexes
(Akkiraju et al. 1995), for the rapid approximation of PH: They use
maxmin as an alternative to random landmark selection, which ensures
that landmarks are locally separated and roughly evenly distributed.
Other uses include sampling points from a computationally intractable
point cloud for the purpose of downstream topological analysis, as when
performing the mapper construction (Singh, Mémoli, and Carlsson 2007);
and the heuristic optimization of a fixed-radius ball cover of a point
cloud, in the sense of minimizing both the number of balls and their
radii (Dłotko 2019).

The maxmin sampling procedure comes with its own limitations in the
analysis of data that vary greatly in density or have many
multiplicities. This is a frequent concern when sparse, heterogeneous,
and incomplete data are modeled as finite pseudometric spaces.
Especially in analyses of medical and healthcare data, underlying
variables can often only be understood as ordinal, and high-dimensional
data sets are commonly analyzed using similarity measures rather than
vector space embeddings. Because measurements are coarse and often
missing, such data also often contain indistinguishable entries---cases
all of whose measurements are equal and that are therefore represented
as multiple instances of the same point. These qualities violate the
assumptions of the ball cover approach and suggest the need for an
ordinal counterpart.

These considerations motivate us to define the \emph{neighborhood
cover}, each set of which may have a different radius but (roughly) the
same cardinality. The two approaches are visually contrasted in
Section\nbs\ref{sec:examples}. Later sections will compare their
performance on several tasks using real-world data, and specifically
data and tasks for which precise sample sizes can be advantageous. From
these experiments, we want to learn both whether cardinality-based
methods outperform distance-based methods (superiority), which might be
expected, but also whether cardinality-based methods are not
outperformed by distance-based methods (non-inferiority), so that they
can be recommended when specific sample sizes are preferable for reasons
other than performance.

\hypertarget{examples}{%
\subsection{Examples}\label{examples}}

\label{sec:examples}

We motivate our alternative sampler using two examples. The first
suggests a practical setting in which fixed-cardinality cover sets are
more desirable for applications. The second provides an abstraction in
which they are better able to detect topological properties.

\begin{example}{Bimodal distribution of risk}

Imagine an intensive care unit whose patients fall roughly into three groups: a large, clinically homogeneous, low-risk majority; a smaller, more heterogeneous, higher-risk group; and a minority of highly distinctive patients who cannot be sorted into either group and are at less predictable risk.
The top panel of Figure\nbs\ref{fig:icu-cover} depicts a simple model of this situation in which each group is represented by a Gaussian distribution.
The points $X$ along the abscissa are sampled randomly from this distribution.

The bottom panel of Figure\nbs\ref{fig:icu-cover} shows how the maxmin and lastfirst procedures generate sequences in $X$ of four landmarks each.
Each procedure begins with a seed landmark $\ell_0$ selected to be reachable by minimum-radius balls (maxmin) or by minimum-cardinality neighborhoods (lastfirst) from all other points in $X$; see the Appendix for details.
Whereas the maxmin landmarks are roughly equally-spaced across the range of $X$, the lastfirst landmarks lie at roughly equal quantiles of $X$.
The lastfirst procedure will be advantageous when subpopulations can be discriminated at different resolutions and the ability to do so is limited primarily by sample size.

\begin{figure}
\includegraphics[width=\textwidth]{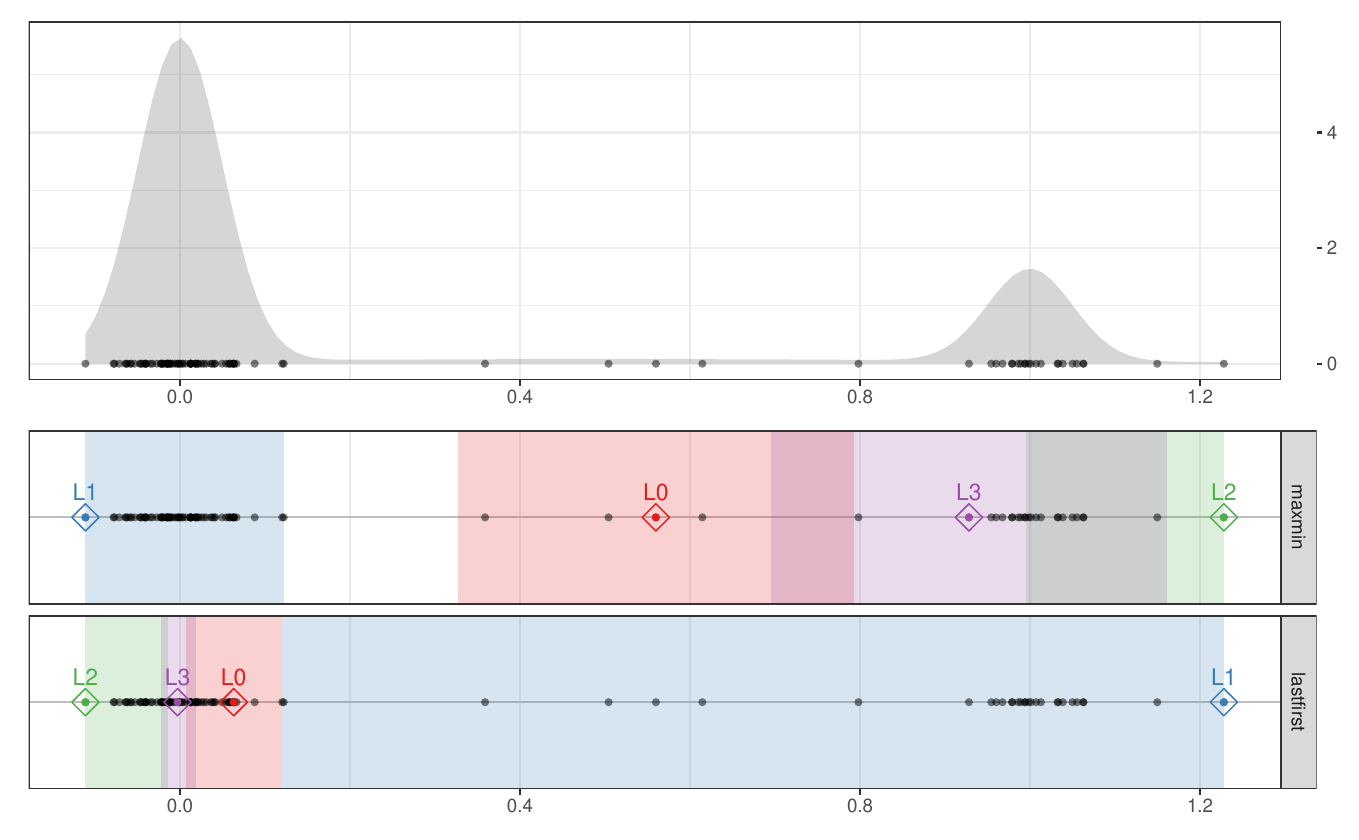}
\caption{
Landmark samples from the imagined ICU sample and their associated covers using two selection procedures.
\label{fig:icu-cover}
}
\end{figure}

\end{example}

\begin{example}{Bimodal density on a circle}
\label{ex:bumpy-circle}

Consider a probability density function that varies significantly over $\Sph^1$ and a sample $X$ from its distribution.
A paradigmatic goal of TDA would be to detect the 1-dimensional feature of $\Sph^1$ from the sample.
Given a small subset of landmarks $L \subset X$, we would want a witness complex $\Wit(L,X)$ to reliably satisfy $H_1(\Wit(L,X)) = 1$.
We would also want to reduce complexity, computational cost, and detection of spurious features.

\begin{figure}
\includegraphics[width=\textwidth]{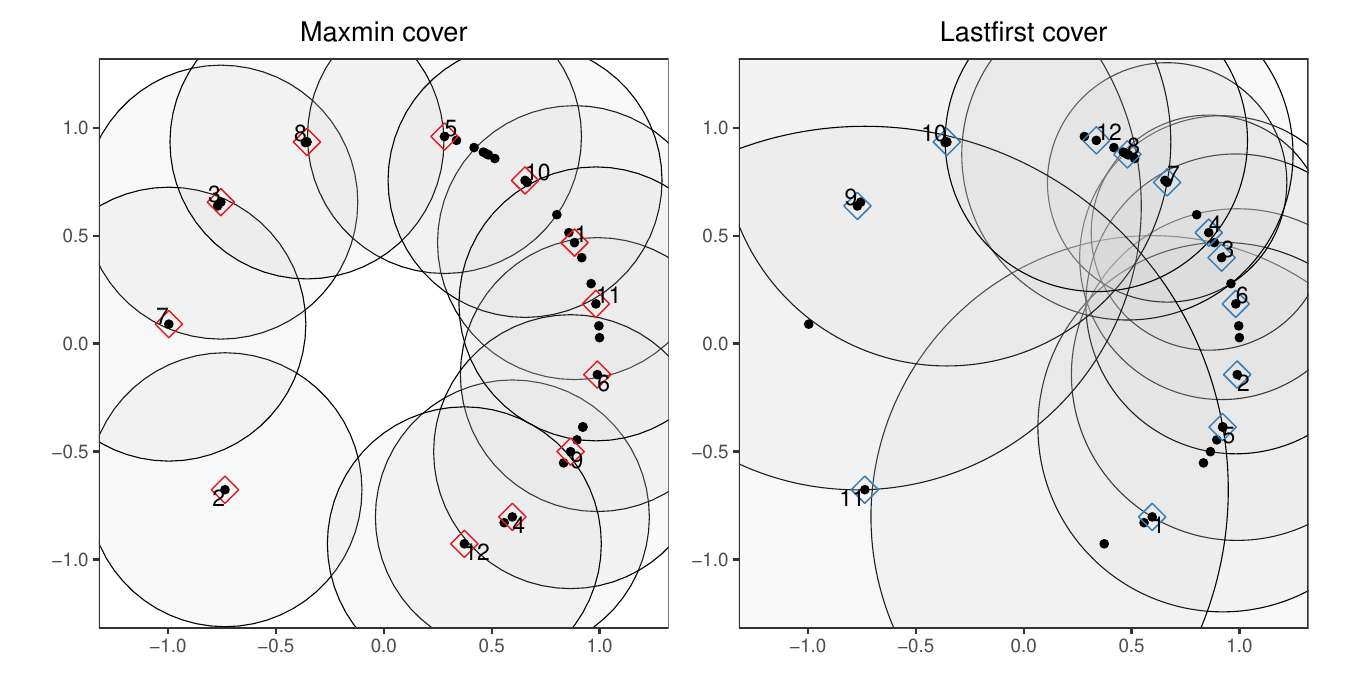}
\caption{
Landmark samples and their covers from the bumpy distribution on the circle using two selection procedures.
\label{fig:bumpy-cover}
}
\end{figure}

Figure\nbs\ref{fig:bumpy-cover} shows the covers obtained from 12 landmarks generated using the maxmin and lastfirst procedures, in both cases increasing the sizes (radii and cardinalities, respectively) of the sets twofold from minimality.
It can be seen that the maxmin witness complex (the nerve of the cover) fails to detect the 1-feature while the lastfirst witness complex succeeds.
As will be shown later in the paper, this superior performance is robust with respect to several parameters governing the distribution and the samplers, in particular that the lastfirst witness complex detects the feature over a wider range of $\lvert L \rvert$.

\end{example}

\hypertarget{materials-and-methods}{%
\section{Materials and Methods}\label{materials-and-methods}}

\label{sec:methods}

This section provides mathematical proofs of algorithms and oher
properties of the landmark samplers (Section\nbs\ref{sec:samplers}),
summarizes their implementation as an \pkg{R} package
(Section\nbs\ref{sec:implementation}), and describes several experiments
performed on real-world data to address our motivating questions
(Sections\nbs\ref{sec:data} and \ref{sec:experiments}).

The \pkg{R} code used to perform experiments, conduct analyses, present
results, and prepare the manuscript are publicly available at the
project repository on GitHub:
\url{https://github.com/corybrunson/lastfirst}.

\hypertarget{samplers}{%
\subsection{Samplers}\label{samplers}}

\label{sec:samplers}

In this section we define both maxmin and its counterpart lastfirst,
provide algorithms for their implementation, and prove some additional
properties about them.

\hypertarget{maxmin-procedure}{%
\subsubsection{Maxmin procedure}\label{maxmin-procedure}}

\label{sec:maxmin}

\begin{definition}[maxmin]\label{def:maxmin}
    Given $(X,d)$ and $Y\subset X$, define the \emph{maxmin set}
    $$\maxmin(Y) = \maxmin(Y;X) = \{ x \in X \wo \cl{Y} \where d(x,Y) = \max_{y\in X\wo \cl{Y}}{d(y,Y)} \}$$
    consisting of \emph{maxmin points}.
\end{definition}

Note that \(\maxmin(Y)\) is nonempty when
\(X \wo \cl{Y} \neq \varnothing\) and that \(\abs{\maxmin(Y)} = 1\) when
\(X\) is in locally general position.

\begin{algorithm}
\caption{Select a maxmin landmark set.}
\label{alg:maxmin}
\begin{algorithmic}[1]
\REQUIRE finite pseudometric space $(X,d)$
\REQUIRE at least one parameter $n \in \bbN$ or $\eps \geq 0$
\REQUIRE seed point $\ell_0 \in X$
\REQUIRE selection procedure $\sigma$
\IF{$\eps$ is not given}
    \STATE $\eps \leftarrow \infty$
\ENDIF
\IF{$n$ is not given}
    \STATE $n \leftarrow 1$
\ENDIF
\STATE $L \leftarrow \varnothing$
\STATE $i \leftarrow 0$
\REPEAT
    \STATE $L \leftarrow L\cup\{\ell_i\}$
    \STATE $i \leftarrow i+1$
    \STATE $\ell_i \leftarrow \sigma(\maxmin(L))$
    \STATE $d_{\max} \leftarrow d(\ell_i,L)$
\UNTIL $d_{\max} < \eps$ and $\abs{L} \geq n$
\RETURN maxmin landmark set $L$
\end{algorithmic}
\end{algorithm}

The \emph{maxmin procedure} for generating a landmark set
\(L\subseteq X\) proceeds as follows (see
Algorithm\nbs\ref{alg:maxmin}). Each step receives a proper subset
\(L\subset X\) and returns a point \(x\in X\wo \cl{L}\).

First, choose a number \(n\leq\uniq{X}\) of landmark points to generate
or a radius \(\eps\geq 0\) for which to require that the balls
\(\{ \cl{B_{\eps}}(\ell) : \ell \in L \}\) cover \(X\). Choose a first
landmark point \(\ell_0\in X\).\footnote{This choice may be arbitrary;
  we consider three selection rules: the first point index in the object
  representing \(X\), selection at random, and the Chebyshev center
  \(\argmin_{x \in X}{ D(x, X \wo \{x\}) }\).} Inductively over
\(i\in\bbN\), if ever \(i\geq n\) or \(d(L,X\wo \cl{L})\leq\eps\), then
stop. Otherwise, when \(L=\{\ell_0,\ldots,\ell_{i-1}\}\), choose
\(\ell_i\in\maxmin(L)\), according to a preferred procedure \(\sigma\)
(see the Appendix). If \(n\) was prescribed, then set
\(\eps=\eps(n)=d(L,X\wo \cl{L})\); if \(\eps\) was prescribed, then set
\(n=n(\eps)=\abs{L}\). Write
\(\mathcal{B}_{n,\eps}(\ell_0)=\{\cl{B_{\eps}}(\ell_i)\}_{i=0}^{n}\) for
the resulting \emph{landmark ball cover of $X$}.

We will write the elements of landmark sets
\(L=\{\ell_0,\ldots,\ell_{n-1}\}\) in the order in which they were
generated. Note that, if \(n=\uniq{X}\) or \(\eps=0\), then
\(\cl{L}=X\). When the procedure stops,
\(X=\bigcup_{i=0}^{n-1}{\cl{B_{\eps}}(\ell_i)}\). This cover is not, in
general, a minimal cover, but it is a minimal landmark cover in the
sense that the removal of \(\cl{B_\eps}(\ell_{n-1})\) or any decrease in
\(\eps\) will yield a collection of sets that fails to cover \(X\). A
``thickened'' cover can be obtained by pre-specifying both \(n\) and
\(\eps\) in such a way that \(n \geq n(\eps)\) and
\(\eps \geq \eps(n)\). In Section\nbs\ref{sec:implementation}, we
describe two adaptive parameters that make these choices easier.

Maxmin is a heuristic, iterative procedure used to select a
well-dispersed subset of points from \(X\), where dispersion is
understood in terms of the interpoint distances of this subset. At each
step, landmarks \(L = \{\ell_0, \ldots, \ell_{i-1}\}\) having been
selected, the next point \(\ell_i\) is selected to maximize its minimum
distance \(d(\ell_j,\ell_i)\) from the \(\ell_j\), \(0 \leq j < i\).
Equivalently, \(\ell_i \in X \wo L\) is selected so that the minimum
radius \(\eps\) required for
\(\ell_i \in \bigcup_{j=0}^{i-1}{\cl{B_\eps}(\ell_j)}\) is maximized.
This is a useful heuristic for constructing a ball cover
\(\mathcal{B} = \{\cl{B_\eps}(\ell_j) \where 0 \leq j < n\}\) centered
at a highly pairwise distant set of landmarks.

We desire in the next section to construct a neighborhood cover whose
centers are analogously dispersed. Accordingly, let us redefine the
maxmin procedure in terms of balls rather than of distances:

\begin{proposition}[maxmin in terms of balls]\label{prop:maxmin}
    Given $(X,d)$ and $Y \subset X$, write $B_\eps(Y) = \bigcup_{y \in Y}{B_\eps(y)}$ and similarly for closed balls, then let
    $$\Eps(Y,X) = \min\{ \eps \where \cl{B_\eps}(Y) = X \}$$
    (using a capital epsilon). Then
    $$\maxmin(Y;X) = X \wo B_{\Epsilon(Y,X)}(Y)$$
\end{proposition}

This yields the alternative loop for Algorithm\nbs\ref{alg:maxmin},
excerpted as Algorithm\nbs\ref{alg:maxmin-ball}.

\begin{algorithm}
\caption{Select a maxmin landmark set, in terms of balls (loop).}
\label{alg:maxmin-ball}
\begin{algorithmic}[1]
\REPEAT
    \STATE $L \leftarrow L\cup\{\ell_i\}$
    \STATE $i \leftarrow i+1$
    \STATE $\eps_{\min} \leftarrow \Eps(L,X)$
    \STATE $F \leftarrow X \wo B_{\eps_{\min}}(L)$ (maxmin set)
    \STATE $\ell_i \leftarrow \sigma(F)$
\UNTIL $\eps_{\min} < \eps$ and $\abs{L} \geq n$
\end{algorithmic}
\end{algorithm}

\hypertarget{lastfirst-procedure}{%
\subsubsection{Lastfirst procedure}\label{lastfirst-procedure}}

\label{sec:lastfirst}

The lastfirst procedure is defined analogously to the maxmin procedure,
substituting nearest neighborhoods, parameterized by their cardinality
\(k\), for balls, parameterized by their radius \(\eps\). Unlike
distance, membership in nearest neighborhoods is not a symmetric
relation: It may be that \(y \in N_k(x)\) while \(x \notin N_k(y)\). We
therefore introduce a companion concept that reverses this relationship:

\begin{definition}[$k$-neighborhoods]
    For $x \in X$, define the \emph{$k$-out-neighborhoods} $N^+_k$ and \emph{$k$-in-neighborhoods} $N^-_k$ of $x$ as the sets
    \begin{align*}
        N^+_k(x) &= \{ y\in X \where y \in N_k(x) \} = N_k(x) \\
        N^-_k(x) &= \{ y\in X \where x \in N_k(y) \}
    \end{align*}
    Given a subset $Y \subseteq X$, we also define
    \begin{align*}
        N^+_k(x,Y) &= Y \cap N^+_k(x) \\
        N^-_k(x,Y) &= Y \cap N^-_k(x)
    \end{align*}
\end{definition}

Note that
\([\{x\}] = N^\pm_0(x) \subseteq \cdots \subseteq N^\pm_{N-1}(x) = X\).
The \(k\)-out-neighborhoods of \(x\) are the \(k\)-nearest neighbors of
\(x\), i.e.~the sets of points in \(X\) that have out-rank at most \(k\)
from \(x\). The \(k\)-in-neighborhoods of \(x\) are the sets of points
in \(X\) from which \(x\) has out-rank at most \(k\).

\begin{example}\label{ex:rank-neighborhoods}
Consider the simple case $X = \{a, b, c, d\}$, visualized below, equipped with the standard Euclidean metric:
\begin{centeredTikz}
    [every label/.append style={text=black!60!blue, font=\scriptsize}]
    \draw[gray] (0,0) -- (5,0);

    \foreach \i in {0,...,3}
        \draw[gray] (\i,0.1) -- + (0,-0.25) node[font=\scriptsize, text=gray, below] {$\i$};
    \draw[gray] (3,0.1) -- + (0,-0.25) node[font=\scriptsize, text=gray, below] {};
    \draw[gray] (5,0.1) -- + (0,-0.25) node[font=\scriptsize, text=gray, below] {$5$};

    \node[circle, draw=blue!60, fill=blue!5, inner sep=0.5mm, label=above:{$a = 1$}] at (1,0) {};
    \node[circle, draw=blue!60, fill=blue!5, inner sep=0.5mm, label=above:{$b = 2$}] at (2,0) {};
    \node[circle, draw=blue!60, fill=blue!5, inner sep=0.5mm, label=above:{$c = 4$}] at (4,0.1) {};
    \node[circle, draw=blue!60, fill=blue!5, inner sep=0.5mm, label=below:{$d = 4$}] at (4,-0.1) {};
\end{centeredTikz}
Compute $N_k^+$ and $N_k^-$ for $a$ and $c$, using $k = 3$:
$$
N_3^+(a) = \{a, b, c, d\},\hspace{1em}
N_3^+(c) = \{b, c, d\},\hspace{1em}
N_3^-(a) = \{a, b\},\hspace{1em}
N_3^-(c) = \{a, b, c, d\}
$$
\end{example}

\begin{remark}
If $\abs{N^\pm_k(x)} = n_k$, then $\abs{N^\pm_k(x,X\wo\{x\})} = n_k - 1$.
\end{remark}

\begin{example}\label{ex:rank-sequence}
Continuing Example\nbs\ref{ex:rank-neighborhoods}, we can compute the other $N_\bullet^+$ and $N_\bullet^-$ for $a$ and $c$:
\begin{align*}
    N_\bullet^+ (a) &= (\abs{N^+_0(a)}, \abs{N^+_1(a)}, \abs{N^+_2(a)}, \abs{N^+_3(a)}) &
    N_\bullet^+ (c) &= (\abs{N^+_0(c)}, \abs{N^+_1(c)}, \abs{N^+_2(c)}, \abs{N^+_3(c)}) \\
    &= (\abs{\{a\}}, \abs{\{a, b\}}, \abs{\{a, b, c, d\}}, \abs{\{a, b, c, d\}}) &
    &= (\abs{\{c, d\}}, \abs{\{c, d\}}, \abs{\{b, c, d\}}, \abs{\{a, b, c, d\}}) \\
    &= (1, 2, 4, 4) &
    &= (2, 2, 3, 4) \\
    \\
    N_\bullet^- (a) &= (\abs{N^-_0(a)}, \abs{N^-_1(a)}, \abs{N^-_2(a)}, \abs{N^-_3(a)}) &
    N_\bullet^- (c) &= (\abs{N^-_0(c)}, \abs{N^-_1(c)}, \abs{N^-_2(c)}, \abs{N^-_3(c)}) \\
    &= (\abs{\{a\}}, \abs{\{a, b\}}, \abs{\{a, b\}}, \abs{\{a, b, c, d\}}) &
    &= (\abs{\{c, d\}}, \abs{\{c, d\}}, \abs{\{a, b, c, d\}}, \abs{\{a, b, c, d\}}) \\
    &= (1, 2, 2, 4) &
    &= (2, 2, 4, 4)
\end{align*}
\end{example}

As with maxmin, a selection procedure \(\sigma\) must be chosen, which
may be needed even if \(X\) is in general position. We operationalize it
as well as our lastfirst procedure by way of a total ordering on the
\(N^\pm_\bullet\).

\begin{definition}[total orders on rank sequences]
    Let $a_n = (a_1,\ldots,a_M),b_n = (b_1,\ldots,b_M)\in\bbN^M$.
    Then $a_n < b_n$ in the reverse lexicographic (revlex) order if $\exists\, i : a_i > b_i \wedge (\forall\, j<i : a_j = b_j)$.
\end{definition}

We impose the revlex order on the \(N_\bullet^+\) and \(N_\bullet^-\).
As explained in the Appendix, this extends the principle behind maxmin
selection---that each subsequent landmark \(\ell_i\) is (among) the last
to be reached by expanding sets about the existing landmarks \(L\)---by
stipulating that, among these candidates, \(\ell_i\) is also (among) the
last to be reached by \emph{at least \(k\) of these sets}, as \(k\)
increases from \(1\).

We now define our counterpart to the maxmin procedure.

\begin{definition}[lastfirst, in terms of neighborhoods]
    Given $(X,d)$ and $Y \subset X$, write $N_k(Y) = \bigcup_{y \in Y}{N_k(y)}$, then let
    $$K(Y,X) = \min\{ k \where N_k(Y) = X \}$$
    Then define the \emph{lastfirst set}
    $$\lf(Y) = \lf(Y;X) = X \wo N_{K(Y,X) - 1}(Y)$$
    consisting of \emph{lastfirst points}.
\end{definition}

The \emph{lastfirst procedure} proceeds analogously to the maxmin
procedure (Algorithm\nbs\ref{alg:lastfirst}). Write
\(\mathcal{N}_{n,k}(\ell_0) = \{ N_k(\ell_i) \}_{i=0}^{n}\) for the
resulting \emph{landmark neighborhood cover of \(X\)}.

\begin{algorithm}
\caption{Calculate the lastfirst landmark sequence from a seed point.}
\label{alg:lastfirst}
\begin{algorithmic}[1]
\REQUIRE finite pseudometric space $(X,d)$
\REQUIRE at least one parameter $n \in \bbN$ or $k \in \bbN$
\REQUIRE seed point $\ell_0 \in X$
\REQUIRE selection procedure $\sigma$
\IF{$k$ is not given}
    \STATE $k \leftarrow \infty$
\ENDIF
\IF{$n$ is not given}
    \STATE $n \leftarrow 1$
\ENDIF
\STATE $L \leftarrow \varnothing$
\STATE $i \leftarrow 0$
\REPEAT
    \STATE $L \leftarrow L\cup\{\ell_i\}$
    \STATE $i \leftarrow i+1$
    \STATE $k_{\min} \leftarrow K(L,X)$
    \STATE $F \leftarrow X \wo N_{k_{\min}-1}(L)$ (lastfirst set)
    \STATE $\ell_i \leftarrow \sigma(F)$
\UNTIL $k_{\min} > k$ and $\abs{L} \geq n$
\RETURN lastfirst landmark set $L$
\end{algorithmic}
\end{algorithm}

We defer further proofs to the Appendix, where we rely on additional
technical definitions. The results apply to a more general conception of
``relative rank'' (our term), one of which is induced by the
pseudometric on any finite pseudometric space but which need only
satisfy the condition that a point is no nearer any other point than
itself.

\begin{corollary}[lastfirst using relative rank]
    Given $Y\subset X$ and a pseudometric $d$ on $X$ with relative rank $q$,
    \begin{align*}
        \lf(Y;X) &= \maxmin(Y;X,q_d) \\
        \lf(X,d) &= \maxmin(X,q_d)
    \end{align*}
\end{corollary}

\begin{proposition}
Algorithm\nbs\ref{alg:lastfirst-landmarks} returns a lastfirst landmark set.
If $n \leq \uniq{X}$ is given as input and $k$ is not, then $\abs{L} = n$.
If $n$ and $k$ are both given, then $\abs{L} \geq n$.
Otherwise, $L$ is minimal in the sense that no proper prefix of $L$ gives a cover of $X$ by $k$-nearest neighborhoods.
\end{proposition}

\begin{example}\label{ex:rank-sequence-order}
    Return again to $X=\{a,b,c,d\}$ from Example\nbs\ref{ex:rank-sequence}. We calculate an exhaustive lastfirst landmark set, seeded with a point of minimal out-neighborhood sequence:
    \begin{enumerate}
        \item We have
        \begin{align*}
            N_\bullet^+(a,\{b,c,d\}) &= (0,1,3,3) &
            N_\bullet^+(b,\{a,c,d\}) &= (0,1,3,3) \\
            N_\bullet^+(c,\{a,b,d\}) &= (1,1,2,3) &
            N_\bullet^+(d,\{a,b,c\}) &= (1,1,2,3)
        \end{align*}
        Under the revlex order, $\argmin_{x\in X}{N_\bullet^+(x,X \wo \cl{\{x\}})}=\{c,d\}$, and we arbitrarily select $\ell_0=c$ and $L=\{c\}$.
        \item For $x \in X \wo \cl{L}$ we now have
        \begin{align*}
            N_\bullet^-(a,\{c\}) &= (0,0,0,1) &
            N_\bullet^-(b,\{c\}) &= (0,0,1,1)
        \end{align*}
        Under the revlex order, $\argmax_{x\in X \wo \cl{\{c\}}}{N_\bullet^-(x,\{c\})}=\{a\}$; we select $\ell_1=a$, so that now $L=\{c,a\}$.
        \item Only one point remains in $X\wo \cl{L}=\{b\}$, so the exhaustive landmark set is $\{c,a,b\}$.
    \end{enumerate}
\end{example}

\hypertarget{implementation}{%
\subsection{Implementation}\label{implementation}}

\label{sec:implementation}

We have implemented both procedures in the \pkg{R} package
\pkg{landmark}\footnote{\url{https://github.com/corybrunson/landmark}},
borrowing a maxmin implementation in the \pkg{Mapper}
package\footnote{\url{https://github.com/peekxc/Mapper}}. Each procedure
is implemented for Euclidean distances in \pkg{C++} using \pkg{Rcpp}
(Eddelbuettel and Francois 2011) and for many other distance metrics and
similarity measures in \pkg{R} using the \pkg{proxy} package.\footnote{\url{https://CRAN.R-project.org/package=proxy}}
For relative rank--based procedures, the user can choose any
tie-handling rule (see the Appendix). The landmark-generating procedures
return the indices of the selected landmarks, optionally together with
the sets of indices of the points in the cover set centered at each
landmark. In addition to the number of landmarks \(n\) and either the
radius \(\eps\) of the balls or the cardinality \(k\) of the
neighborhoods, the user may also specify multiplicative and additive
extension factors for \(n\) and for \(\eps\) or \(k\). These will
produce additional landmarks (\(n\)) and larger cover sets (\(\eps\) or
\(k\)) with increased overlaps, in order to construct more overlapping
covers. A multiplicative factor \(a \geq 0\) extends the radius
(cardinality) of each ball (neighborhood) in a maxmin (lastfirst) cover
by a factor of \(a\) (so that \(a = 0\) produces no change), while an
additive factor \(b \geq 0\) extends the radius (cardinality) by \(b\)
units.

For the bumpy circle experiments we also used the \pkg{simplextree}
package\footnote{\url{https://CRAN.R-project.org/package=simplextree}}
and \pkg{Python GUDHI} (Maria et al. 2014) by way of the
\pkg{reticulate} and \pkg{interplex} packages.\footnote{\url{https://CRAN.R-project.org/package=reticulate}}\footnote{\url{https://CRAN.R-project.org/package=interplex}}

\hypertarget{validation}{%
\subsubsection{Validation}\label{validation}}

We validated the maxmin and lastfirst procedures against manual
calculations on several small example data sets, including that of
Example\nbs\ref{ex:relative-rank}. We also validated the \pkg{C++} and
\pkg{R} implementations against each other on several larger data sets,
including as part of the benchmark tests reported in the next section.

\hypertarget{benchmark-tests}{%
\subsubsection{Benchmark tests}\label{benchmark-tests}}

We benchmarked the \pkg{C++} and \pkg{R} implementations on three data
sets: uniform samples from the unit circle \(\Sph^1\subset\bbR^2\)
convoluted with Gaussian noise, samples with duplication from the
integer lattice \([0,23]\times[0,11]\) using the probability mass
function \(p(a,b) \propto 2^{-ab}\), and patients recorded at each
critical care unit in MIMIC-III using RT-transformed data and cosine
similarity (Section\nbs\ref{sec:data}). We conducted benchmarks using
the \pkg{bench} package (Hester 2020) on the University of Florida
high-performance cluster HiPerGator.

\hypertarget{simulations-and-data}{%
\subsection{Simulations and data}\label{simulations-and-data}}

\label{sec:data}

The experiments described in Section\nbs\ref{sec:experiments} make use
of the sample simulations and real-world data sets detailed here.

\hypertarget{bumpy-circle}{%
\subsubsection{Bumpy circle}\label{bumpy-circle}}

As noted in Example\nbs\ref{ex:bumpy-circle}, the most basic tools to
detect homological features in data may fail when the data are not
evenly distributed across the space of interest. Much of the TDA
literature is geared toward addressing this problem. The lastfirst
procedure was designed in part as a deterministic sampler analogous to
maxmin that is robust to such variation in density. We use data sets
sample from a parameterized family of distributions to test its
performance.

The general idea is that the data used as input for the landmark
proceduces will be comprised of sparsely sampled points on \(\Sph^1\)
that has two regions of higher density. We identify \(\Sph^1\) with
\(\bbR / 2\pi\bbR\) via the parameterization
\(t \mapsto (\cos(2\pi t), \sin(2\pi t))\). The density function will be
a weighted mixture distribution of three distributions: a uniform
distribution \(U(0,2\pi)\) on \(\Sph^1\) and two Gaussian distributions
\(N(0,\sigma^2)\) and \(N(\mu,\sigma^2)\) on \(\bbR\) taken modulo
\(2\pi\). In addition to the sample size, we varied the mixture weights
\(w_0,w_1,w_2\) and the spreads and relative positions of the Gaussians.
We also varied the number of landmarks sampled using each procedure and
the multiplicative and additive extensions to the cover sets obtained.
The values taken by each parameter are listed in
Table\nbs\ref{tab:bumpy-grid}, and we performed one experiment for each
combination.

\begin{table}\label{tab:bumpy-grid}
\centering
\begin{tabular}{|l|c|} \hline
Parameter & Values \\ \hline
sample size & $n = 12, 36, 60$ \\
weight of uniform & $w_0 = 0, 0.01, 0.05, 0.1$ \\
angular distance & $\mu_2 = \pi, \frac{3}{4}\pi, \frac{1}{2}\pi$ \\
standard deviation & $\sigma = \frac{1}{6}\pi, \frac{1}{3}\pi, \frac{1}{2}\pi$ \\
relative weights (Gaussians) & $r = w_1 / w_2 = 1, 3, 10$ \\
sampling procedure & maxmin, lastfirst \\
multiplicative extension & $\operatorname{ext}_\times = 0, 1, 2$ \\
additive extension & $\operatorname{ext}_+ = 0, .1, .2$ (maxmin) \\
 & $\operatorname{ext}_+ = 0, n / 12, n / 6$ (lastfirst) \\ \hline
\end{tabular}
\caption{Parameter values used in bumpy circle simulations.}
\end{table}

\hypertarget{mimic-iii}{%
\subsubsection{MIMIC-III}\label{mimic-iii}}

\label{sec:mimic}

The open-access critical care database MIMIC-III (``Medical Information
Mart for Intensive Care''), derived from the administrative and clinical
records for 58,976 admissions of 46,520 patients over 12 years and
maintained by the MIT Laboratory for Computational Physiology and
collaborating groups, has been widely used for education and research
(Goldberger et al. 2000; A. E. W. Johnson et al. 2016; A. Johnson,
Pollard, and Mark 2016). Researchers may gain access to the database by
completing a training course in human subjects research and signing a
data use agreement.\footnote{\url{https://mimic.mit.edu/docs/gettingstarted/}}
For our analyses we included data for patients admitted to five care
units: coronary care (CCU), cardiac surgery recovery (CSRU), medical
intensive care (MICU), surgical intensive care (SICU), and
trauma/surgical intensive care (TSICU).\footnote{\url{https://mimic.physionet.org/mimictables/transfers/}}

We ran experiments on these data using three distance measures defined
on patient admissions: First, we extracted or calculated the vital sign,
laboratory test, administrative, procedural, and demographic variables
used by Lee, Maslove, and Dubin (2015). From these, we calculated the
cosine similarity (simplified from its form there, using one-hot
encoding of categorical variables) and the Gower distance (Gower 1971).
As a domain-agnostic alternative, we extracted the set of ICD-9/10 codes
from the patient's record and several categorical demographic variables:
age group (18--29, decades 30--39 through 70--79, and 80+), recorded
gender (M or F), stated ethnicity (41 values),\footnote{White,
  Black/African American, Unknown/Not Specified, Hispanic or Latino,
  Other, Unable to Obtain, Asian, Patient Declined to Answer, Asian --
  Chinese, Hispanic Latino -- Puerto Rican, Black/Cape Verdean, White --
  Russian, Multi Race Ethnicity, Black/Haitian, Hispanic/Latino --
  Dominican, White -- Other European, Asian -- Asian Indian, Portuguese,
  White -- Brazilian, Asian -- Vietnamese, Black/African, Middle
  Eastern, Hispanic/Latino -- Guatemalan, Hispanic/Latino -- Cuban,
  Asian -- Filipino, White -- Eastern European, American Indian/Alaska
  Native, Hispanic/Latino -- Salvadoran, Asian -- Cambodian, Native
  Hawaiian or Other Pacific Islander, Asian -- Korean, Asian -- Other,
  Hispanic/Latino -- Mexican, Hispanic/Latino -- Central American
  (Other), Hispanic/Latino -- Colombian, Caribbean Island, South
  American, Asian -- Japanese, Hispanic/Latino -- Honduran, Asian --
  Thai, American Indian/Alaska Native Federally Recognized Tribe} stated
religion,\footnote{Catholic, unspecified/unobtainable/missing,
  Protestant Quaker, Jewish, other, Episcopalian, Greek Orthodox,
  Christian Scientist, Buddhist, Muslim, Jehovah's Witness,
  Unitarian-Universalist, 7th Day Adventist, Hindu, Romanian Eastern
  Orthodox, Baptist, Hebrew, Methodist, Lutheran} marital
status\footnote{married, single, widowed, divorced, unknown/missing,
  separated, life partner}, and type of medical insurance\footnote{Medicare,
  private, Medicaid, povernment, self pay}. Following Zhong, Loukides,
and Gwadera (2020), we transformed these \emph{relational-transaction
(RT)} data into a binary case-by-variable matrix
\(X \in \bbB^{n \times p}\) and again calculated cosine similarity. In
both cases, cosine was converted to a distance measure by subtraction
from 1. Because cosine similarity is monotonically related to the angle
metric, our topological results will be the same as if we had used the
latter.

\hypertarget{mexican-department-of-health}{%
\subsubsection{Mexican Department of
Health}\label{mexican-department-of-health}}

The Mexican Department of Health (MXDH) has released official
open-access data containing an assortment of patient-level clinical
variables related to COVID-19 infection and outcomes. These data have
been compiled into a database and made freely available on
Kaggle\footnote{\url{https://www.kaggle.com/lalish99/covid19-mx}}, a
collaborative data science platform. The data we obtained includes
information regarding over 724,000 patients confirmed to be
COVID-positive via diagnostic laboratory testing. Two main types of
information are present for each patient: (1) dates, and (2) categorical
or binary variables. The former are dates associated with key moments in
the clinical course of infection such as symptom onset, admission to a
healthcare institution, and death (if applicable). The categorical and
binary fields encode clinical factors likely to be associated with
COVID-19 infection, severity, or outcome. These variables include
information such as sex, state of patient residence, and intubation
status, as well as binary fields encoding the presence or absence of a
wide variety of comorbidities such as asthma, hypertension,
cardiovascular disease. Though these variables are categorical rather
than continuous/numeric, there are sufficiently many of them
(\(\approx 50\)) to potentially distinguish between many patient
phenotypes. Further, this data set is very complete in that every
patient is required to contain a valid value for every field, which
minimizes concerns around missing data.

\hypertarget{experiments}{%
\subsection{Experiments}\label{experiments}}

\label{sec:experiments}

\hypertarget{bumpy-circle-1}{%
\subsubsection{Bumpy circle}\label{bumpy-circle-1}}

The goal of these experiments is to find and illustrate a specific
example of a situation where the lastfirst covering algorithm
outperforms its maxmin counterpart. We would expect such a situation to
occur when density within the input data is highly variable, motivating
the choice of experimental data sets described above. Differences
between maxmin and lastfirst can be illustrated by comparing the
persistence of relevant topological features.

For this example, we generated maxmin ball covers
\(\mathcal{B}_{m,\eps_m}(\ell_0)\) and lastfirst neighborhood covers
\(\mathcal{N}_{m,k_m}(\ell_0)\) for each \(m = 1,\ldots,n' \leq n\) and
each sample, in all cases starting with a random seed point \(\ell_0\).
We computed the \emph{landmark persistence} of the known 0- and
1-dimensional features of \(\Sph^1\)---a single connected component and
a single loop---which we define to be the range of numbers of landmarks
over which the landmark cover recovered each feature. We report the
shared ranges over which both features were recovered, which we term the
\emph{landmark dominance} after de Silva and Carlsson (2004), though the
results differ only slightly from the ranges over which the loop was
recovered. These we obtained by computing the sequence of Betti numbers
\(\beta_{i,m} = \beta_i(\mathcal{B}_{m,\eps_m}(\ell_0))\) or
\(\beta_{i,m} = \beta_i(\mathcal{N}_{m,k_m}(\ell_0))\) of the nerves of
the associated sequence of covers.

\begin{remark}
Our cover sets are taken to lie in $\Sph^1$, though we computed and visualized them in $\bbR^2$.
Working in $\bbR^2$, if $p \in \Sph^1$ and $d > 0$, then $B_d(p) \cap \Sph^1$ is either an interval of $\Sph^1$ or $\Sph^1$ itself, so in topological terms the difference does not matter. For the same reason, no spurious 1-features are possible.
If we were to report persistence in radii, we would need to transform Euclidean distance in $\bbR^2$ to angular distance in $\Sph^1$; but we don't, so we didn't.
\end{remark}

\hypertarget{covers-and-nerves}{%
\subsubsection{Covers and nerves}\label{covers-and-nerves}}

Cardinality reduction techniques can be used to model a large number of
cases represented by a large number of variables as a smaller number of
clusters with similarity or overlap relations among them. The
deterministic maxmin and lastfirst procedures provide \emph{fuzzy}
clusters---that is to say, clusters that allow for some overlap,
equivalently the sets of a cover, in contrast to \emph{crisp} partitions
that do not overlap---defined by proximity to the landmark cases and
relations defined by their overlap. The clusters obtained by these
procedures occupy a middle ground between the regular intervals or
quantiles commonly used to cover samples from Euclidean space and the
emergent clusters obtained heuristically by penalizing between-cluster
similarity and rewarding within-cluster similarity. The maxmin procedure
produces cover sets of fixed radius, analogous to overlapping intervals
of fixed length, while the lastfirst procedure produces cover sets of
(approximately) fixed size, analogous to the quantiles of an adaptive
cover. This makes them natural solutions to the task of covering an
arbitrary finite metric space that may or may not contain important
geometric or topological structure (Singh, Mémoli, and Carlsson 2007).

As a practical test of this potential, we loosely followed the approach
of Dłotko (2019) to construct covers and their nerves for each care unit
of MIMIC-III, using maxmin and lastfirst. We varied the number of
landmarks (6, 12, 24, 36, 48, 60, 120) and the multiplicative extension
of the cover sets' sizes (0, .1, .2). We evaluated the procedures in
three ways:

\begin{itemize}
\tightlist
\item
  \textbf{Clustering quality:} While clustering quality measures might
  be useful, most, including almost all that have been proposed for
  fuzzy clusterings, rely on coordinate-wise calculations, specifically
  data and cluster centroids (Bouguessa, Wang, and Sun 2006; Wang and
  Zhang 2007; Falasconi et al. 2010). Coordinates are not assumed in the
  general setting of finite pseudometric spaces and are unavailable or
  not meaningful in many real-world settings. To our knowledge, the sole
  exception to have appeared in a comprehensive comparison of such
  measures is the \emph{modified partition coefficient} (Dave 1996),
  defined as
  \[\operatorname{MPC}=1-\frac{k}{k-1}(1-\frac{1}{n}\sum_{i=1}^{n}{\sum_{j=1}^{k}{{u_{ij}}^2}})\]
  where \(U=(u_{ij})\) is the \(n\times k\) fuzzy partition matrix:
  \(u_{ij}\) encodes the extent of membership of point \(x_i\) in
  cluster \(c_j\), and \(\sum_{j=1}^{k}{u_{ij}}=1\) for all \(i\). When
  a point \(x_i\) is contained in \(m\) cover sets \(c_j\), we equally
  distribute its membership so that \(u_{ij}=\frac{1}{m}\) when
  \(x_i\in c_j\) and \(u_{ij}=0\) otherwise. Thus, the MPC quantifies
  the extent of overlap between all pairs of clusters. Like the
  partition coefficient from which it is adapted, the MPC takes the
  value \(1\) on crisp partitions and is penalized by membership
  sharing, but it is standardized so that its range does not depend on
  \(k\).
\item
  \textbf{Discrimination of risk:} For purposes of clinical phenotyping,
  patient clusters are more useful that better discriminate between low-
  and high-risk subgroups. We calculate a cover-based risk estimate from
  individual outcomes \(y_i\) as follows: For each cover set
  \(c_j\subset X\), let
  \(p_j=\frac{1}{\abs{c_j}}\sum_{x_i\in c_j}{y_i}\) be the incidence of
  the outcome in that set. Then compute the weighted sum
  \(q_i=\sum_{x_i\in c_j}{u_{ij}p_j}\) of these incidence rates for each
  case. We measure how well the cover discriminates risk as the area
  under the receiver operating characteristic curve (AUROC).
\end{itemize}

We hypothesized that lastfirst covers would exhibit less overlap than
maxmin covers by virtue of their greater sensitivity to local density,
and that they would outperform maxmin covers at risk prediction by
reducing the sizes of cover sets in denser regions of the data (taking
advantage of more homogeneous patient cohorts).

\hypertarget{interpolative-nearest-neighbors-prediction}{%
\subsubsection{Interpolative nearest neighbors
prediction}\label{interpolative-nearest-neighbors-prediction}}

Landmark points may also be used to trade accuracy for memory in
neighborhood-based prediction modeling. Consider the following approach:
A modeling process involves predictor data \(X \in \bbR^{n \times p}\)
and response data \(y \in \bbR^{n \times 1}\), partitioned into training
and testing sets \(X_0,X_1\) and \(y_0,y_1\) according to a partition
\(I_0 \sqcup I_1 = \{1,\ldots,n\}\) of the index set. Given
\(x \in X_1\), a nearest neighbors model computes the prediction
\(p(x) = \frac{1}{k}\sum_{q(x,x_i) \leq k}{y_i}\) by averaging the
responses for the \(k^\text{th}\) nearest neighbors of \(x\) in \(X_0\).
By selecting a landmark set \(L \subset X_0\), a researcher can reduce
the computational cost of the model as follows: For each \(\ell \in L\),
calculate \(p(\ell)\) as above. Then, for each \(x \in X_1\), calculate
\(p_L(x) = \sum_{\ell \in L}{w(d(x,\ell)) p(\ell)} / \sum_{\ell \in L}{w(d(x,\ell))}\),
where \(w : \bbR_{\geq 0} \to \bbR_{\geq 0}\) is a weighting function
(for example, \(w(d)=d^{-1}\)). The nearest neighbor predictions for
\(L\) thus serve as proxies for the responses associated with \(X_0\).
We refer to these as interpolative nearest neighbors (INN) models.

We took this approach to the prediction of in-hospital mortality for
patients with records in each critical care unit of MIMIC-III. We then
implemented the following procedure:

\begin{enumerate}
\def\labelenumi{\arabic{enumi}.}
\tightlist
\item
  Determine a nested \(6 \times 6\)--fold split for train--tune--test
  cross-validation. That is, partition
  \([n] = \bigsqcup_{i=1}^{6}{I_i}\) into roughly equal parts, and
  partition each \([n] \wo \cl{I_i} = \bigsqcup_{j=1}^{6}{J_{ij}}\) into
  roughly equal parts.
\item
  Iterate the following steps over each \(i,j\):

  \begin{enumerate}
  \def\labelenumii{\alph{enumii})}
  \tightlist
  \item
    Generate a sequence \(L\) of landmarks from the points
    \(X_{([n] \wo \cl{I_i}) \wo \cl{J_{ij}}}\).
  \item
    Identify the \(180\) nearest neighbors \(N^+_{180}(\ell)\) of each
    landmark \(\ell\). This was a fixed parameter, chosen for being
    slightly larger than the optimal neighborhood size in a previous
    study of individualized models (Lee, Maslove, and Dubin 2015).
  \item
    Find the value of \(k \in [180]\) and the weighting function \(w\)
    (among those available) for which the predictions
    \(p_L : X_{J_{ij}} \to [0,1]\) maximize the AUROC.
  \item
    Use the AUROC to evaluate the performance of the predictions
    \(p_L : X_{I_i} \to [0,1]\) using these \(k\) and \(w\).
  \end{enumerate}
\end{enumerate}

We replicated the experiment for each combination of procedure (random,
maxmin, lastfirst), number of landmarks (\(\abs{L}=36,60,180,360\)), and
each of several weighting functions (integer rank, triangle, inverse
distance, Gaussian). We hypothesized that, as measured by overall
accuracy of the resulting predictive model, the maxmin and lastfirst
procedures would outperform random selection, and that lastfirst would
outperform maxmin, for similar reasons to those in the previous section.

\hypertarget{results}{%
\section{Results}\label{results}}

\label{sec:results}

\hypertarget{definition}{%
\subsection{Definition}\label{definition}}

We introduce the following definitions in Section\nbs\ref{sec:methods}:

\begin{definition}[sampling procedures in terms of cover sets]\label{def:maxmin-lastfirst}
    Given $(X,d)$ and $Y \subset X$, write
    \begin{align*}
    \cl{B_\eps(Y)} &= \bigcup_{y \in Y}{\cl{B_\eps(y)}} \\
    N_k(Y) &= \bigcup_{y \in Y}{N_k(y)}
    \end{align*}
    Take
    \begin{align*}
    \Eps(Y,X) &= \min\{ \eps \where \cl{B_\eps(Y)} = X \} \\
    K(Y,X) &= \min\{ k \where N_k(Y) = X \}
    \end{align*}
    Then define the \emph{maxmin} and \emph{lastfirst sets}
    \begin{align*}
    \maxmin(Y) = \maxmin(Y;X) &= X \wo B_{\Epsilon(Y,X)}(Y) \\
    \lf(Y) = \lf(Y;X) &= X \wo N_{K(Y,X) - 1}(Y)
    \end{align*}
    consisting of \emph{maxmin} and \emph{lastfirst points}, respectively.
\end{definition}

As suggested by their definitions, these procedures arise from the
construction of conditionally minimal covers of \(X\), where minimality
is defined in terms of the common radius (maxmin) or cardinality
(lastfirst) of the sets. The centers of the cover sets comprise the
landmark points, beginning from an arbitrarily selected first landmark.
In practice, we suggest a Chebyshev center
\(\argmin_{x \in X}{ \min\{ \eps : \cl{B_\eps(x)} = X \} }\) as a
starting landmark for the maxmin procedure, in that it provides an
(unconditionally) minimal one-set cover.\footnote{An analogously defined
  point \(\argmin_{x \in X}{ \min\{ k : N_k(x) = X \} }\) might be used
  for the lastfirst procedure; when \(X\) is not in locally general
  position, it locates a landmark that is equidistant to as many
  farthest neighbors as possible, i.e.~the center of a maximally
  populated circumcenter of \(X\).}

\hypertarget{implementation-1}{%
\subsection{Implementation}\label{implementation-1}}

The definition of our lastfirst procedure is analogous to that of
maxmin, substituting ranks in the role of distances. In this way,
lastfirst is an alternative to maxmin that is adaptive to the local
density of the data, similar to the use of fixed quantiles in place of
fixed-length intervals. The maxmin and lastfirst procedures implicitly
construct a minimal cover whose sets are centered at the selected
landmarks, and the fixed-radius balls of maxmin correspond to the
fixed-cardinality neighborhoods of lastfirst. The rank-based procedures
are more combinatorially complex and computationally expensive,
primarily because relative ranks are asymmetric, which doubles (in the
best case) or squares (in the worst case) the number of distances that
must be calculated. Nevertheless, the procedure can be performed in a
reasonable time for many real-world uses.

\hypertarget{illustration}{%
\subsubsection{Illustration}\label{illustration}}

Consider the ``necklace'' data set adapted from Yoon and Ghrist (2020).
The points are sampled in the plane from a large, low-density circular
region (the ``string'') and from several smaller, higher-density
circular regions (the ``beads'') evenly spaced along the string. Only a
method of detecting topology that is adaptive to the different scales of
the string and of the beads will recover both the large feature and the
smaller features of dimension 1. We use small landmark samples to
illustrate the differences between the maxmin and lastfirst selection
procedures. Publicly available software tools are not yet available to
compute persistent homology for arbitrary sequences of simplicial maps,
but we suggest that the relative strengths of maxmin and lastfirst would
be revealed by a comparison of the persistent features obtained by
sequences of covers obtained by both procedures on this data set.

\begin{figure}
\includegraphics[width=\textwidth]{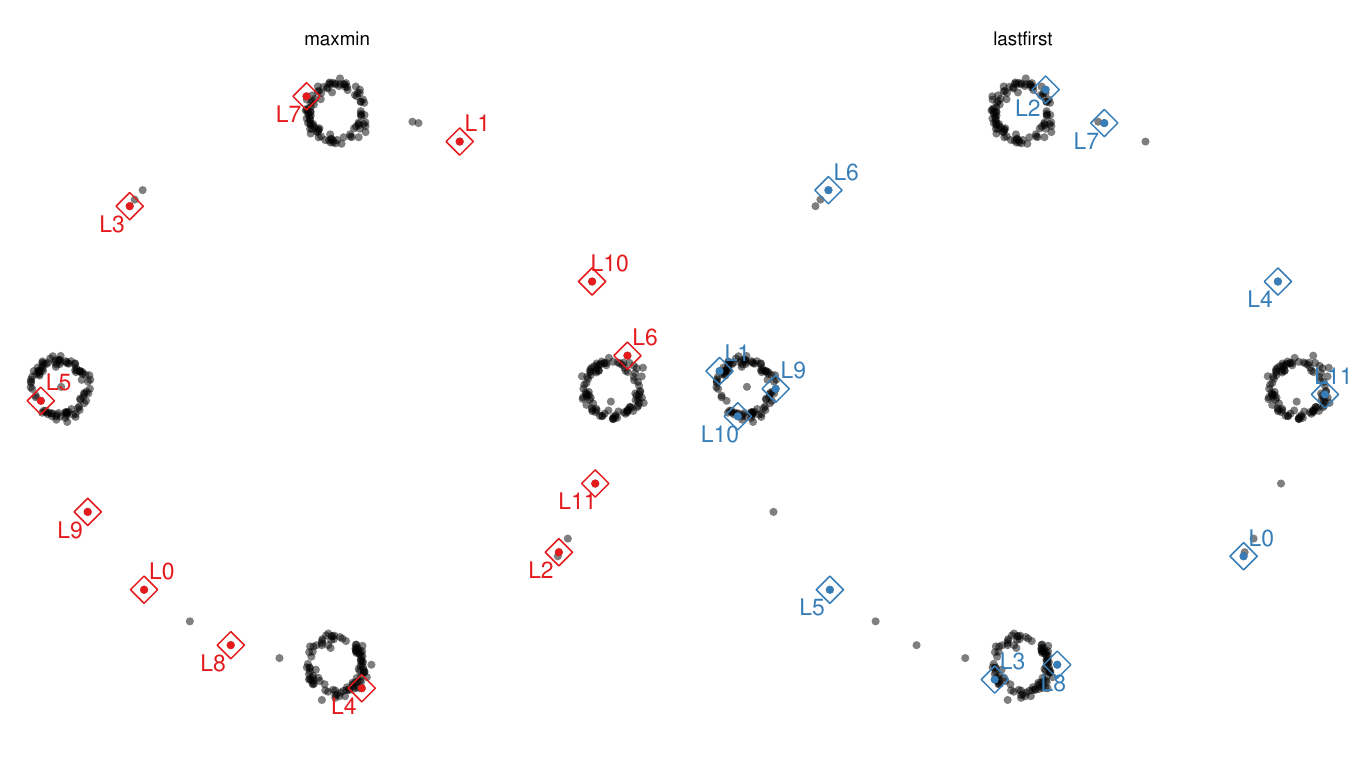}
\caption{
Landmark samples of size 12 from a necklace data set using two selection procedures.
\label{fig:necklace}
}
\end{figure}

\hypertarget{benchmark-tests-1}{%
\subsubsection{Benchmark tests}\label{benchmark-tests-1}}

\begin{figure}
\includegraphics[width=.6666667\textwidth]{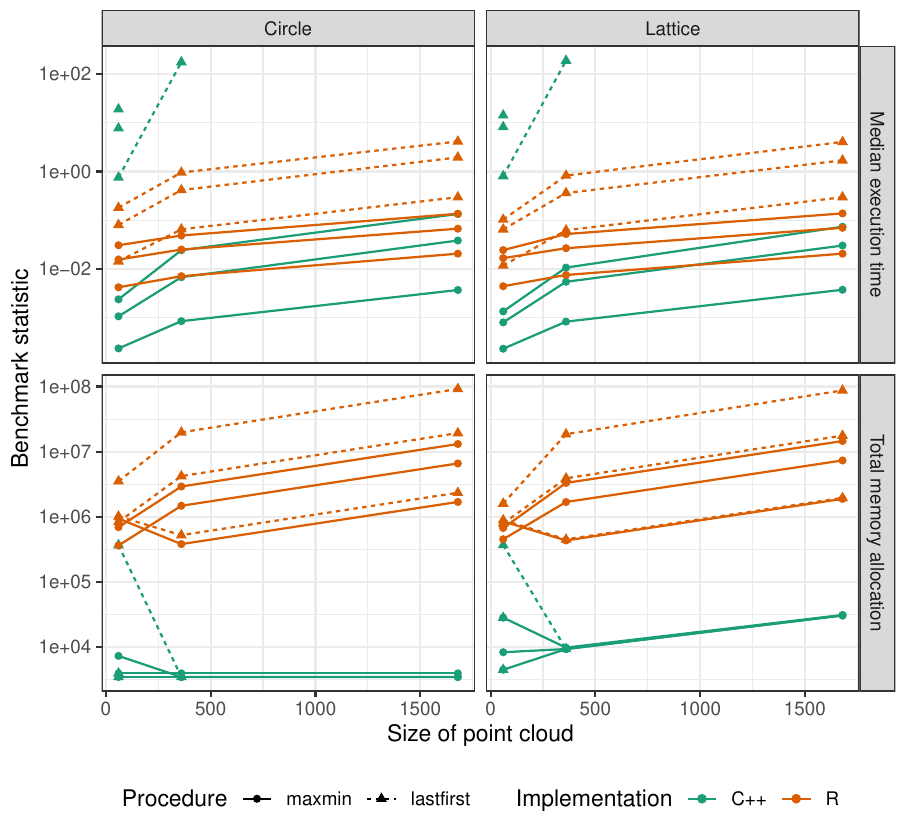}
\includegraphics[width=.3333333\textwidth]{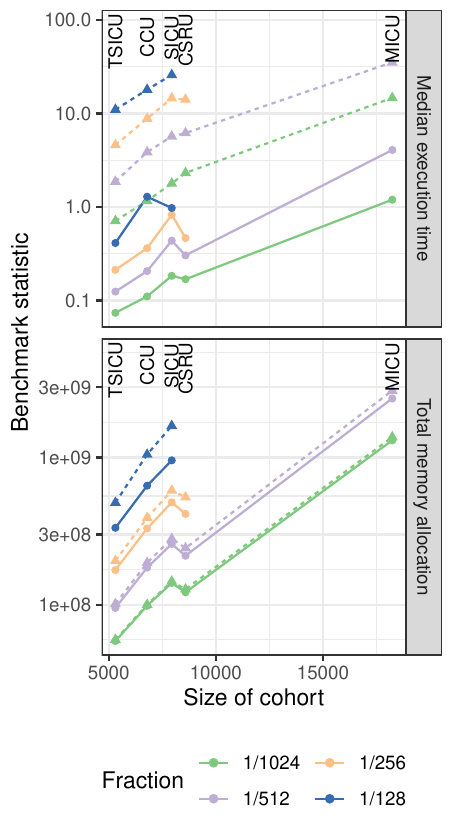}
\caption{
Benchmark results for computing landmarks on two families of artificial data (circle and lattice) and one collection of empirical data (RT-similarity space of critical care units in MIMIC-III). Some points are missing because benchmark tests did not complete within 1 hour.
\label{fig:benchmark}
}
\end{figure}

Benchmark results are reported in Figure\nbs\ref{fig:benchmark}. The
\pkg{R} implementation of maxmin used orders of magnitude more memory
and took slightly longer than the \pkg{C++} implementation. They
appeared to scale slightly better in terms of time than the lastfirst
implementations. The additional calculations required for the lastfirst
procedure increase runtimes by a median factor of 2.5 in our \pkg{R}
implementations. The \pkg{C++} implementation of lastfirst is based on
combinatorial definitions and not optimized for speed, and as a result
took much longer and failed to complete in many of our tests.

\hypertarget{experiments-1}{%
\subsection{Experiments}\label{experiments-1}}

\hypertarget{bumpy-circle-2}{%
\subsubsection{Bumpy circle}\label{bumpy-circle-2}}

\begin{figure}
\includegraphics[width=\textwidth]{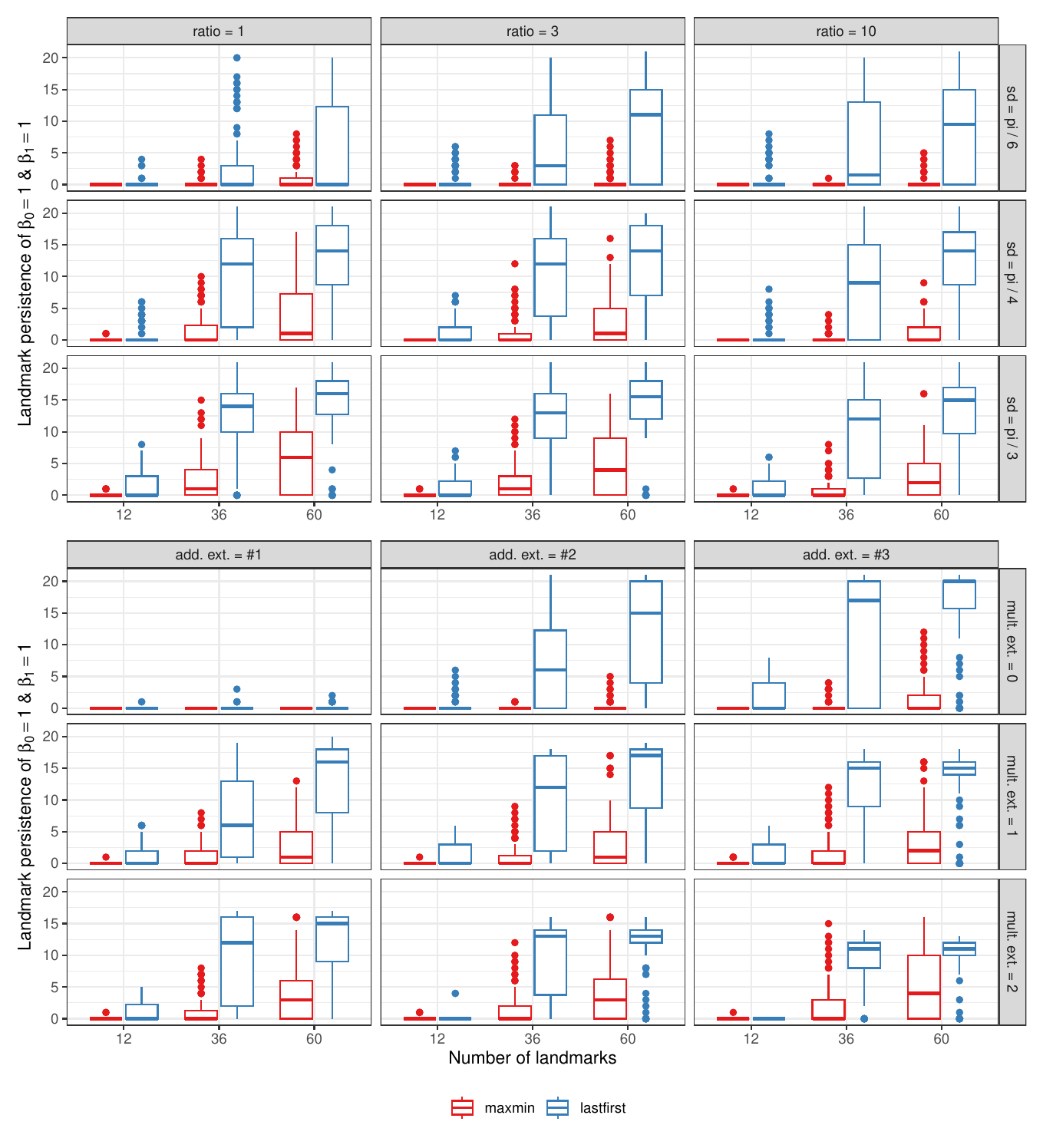}
\caption{
Landmark persistence of maxmin and lastfirst covers, faceted by density parameters (above) and by multiplicative and additive extensions (below).
See Section\nbs\ref{sec:implementation} for details.
The standard deviation is that of the two Gaussian distributions and the ratio is that between their mixture weights.
The additive extensions applied to ball radii in the case of maxmin and to neighborhood cardinalities in the case of lastfirst.
\label{fig:bumpy-distribution}
}
\end{figure}

Figures\nbs\ref{fig:bumpy-distribution} compares the landmark dominance
of the Betti numbers \(\beta_0 = \beta_1 = 1\) of \(\Sph^1\) using
maxmin and lastfirst covers, respectively faceted by two parameters
governing the density function on \(\Sph^1\) and by the two extension
parameters used to construct the covers. Both plots depict all
experimental results. Boxes are bounded by quartiles and whiskers by 1.5
interquartile ranges.

The top plot shows that landmark dominance was greater, using both
procedures, when the Gaussian distributions were wider. This is
consistent with the general pattern of more uniform distributions on or
near manifolds leading to better detection of homological features. In
contrast, the relative numbers of samples from the two Gaussians had
little impact on landmark dominance. (The proportion of uniformly
sampled points and the angle between the Gaussians' centers had
similarly minor effects.)

The bottom plot demonstrates the importance of cover set extension to
the successful detection of homology. Both procedures obtained the
features of \(\Sph^1\) across much wider ranges of landmarks when
multiplicative or additive extensions were used, and for the most part
the improvement due to both extensions was complementary. The extensions
are not interchangeable: Increasing the ball radii by a fixed multiple
has a smaller effect in sparser regions than increasing the neighborhood
cardinalities by the same multiple, and a greater multiplicative
extension might result in similar performance. This would itself still
support the basic lesson that a cardinality-based approach outperforms
the analogous radius-based one. Neither procedure reliably detected the
homology of \(\Sph^1\) without cover extensions, but it is worth noting
that only the lastfirst witness complexes were able to detect it at all.

\hypertarget{covers-and-nerves-1}{%
\subsubsection{Covers and nerves}\label{covers-and-nerves-1}}

\begin{figure}
\includegraphics[width=.5\textwidth]{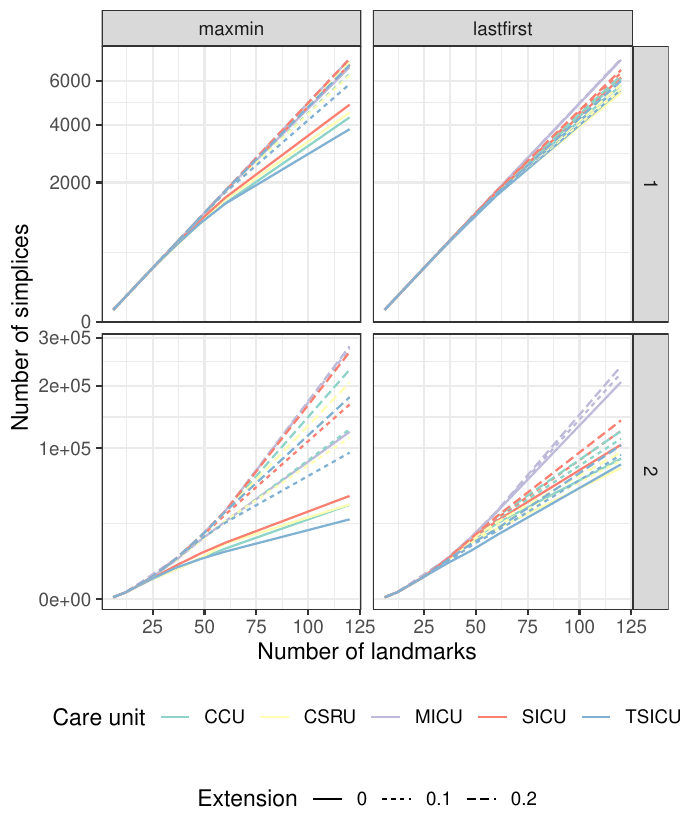}
\includegraphics[width=.5\textwidth]{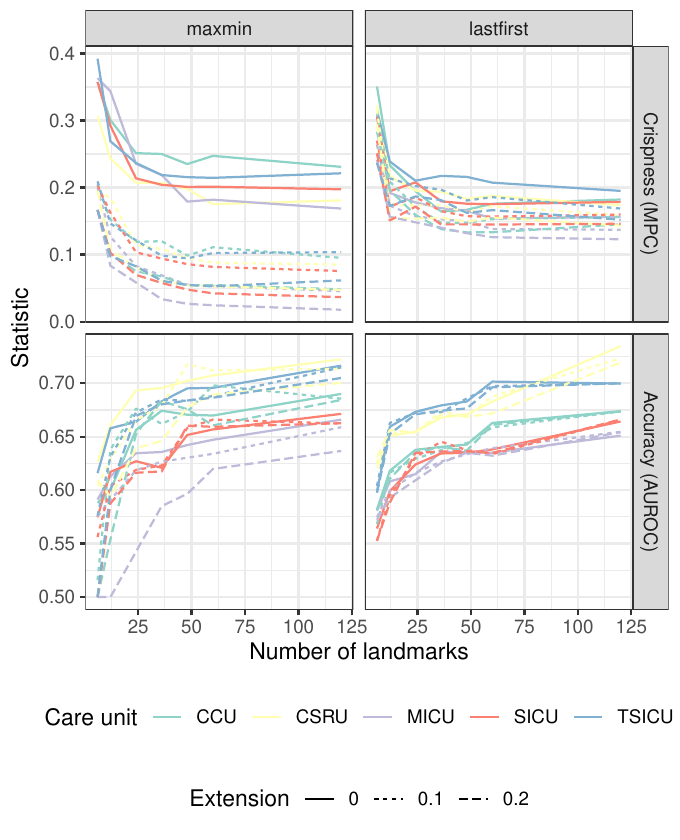}
\caption{
Summary and evaluation statistics versus number of 0-simplices (landmarks) for the covers generated using the maxmin and lastfirst procedures, with three multiplicative extensions in their size.
Left: the sizes of their nerves, as numbers of 1- and 2-simplices, using a square root--transformed vertical scale.
Right: the modified partition coefficient (MPC) and the c-statistic of the risk prediction model based on the cover sets (AUROC).
\label{fig:cover-mimic}
}
\end{figure}

Figure\nbs\ref{fig:cover-mimic} presents, for the analysis of the five
MIMIC-III care units, the sizes of the nerves of the covers and the two
evaluation statistics as functions of the number of landmarks. The
numbers of 1- and of 2-simplices grew at most roughly quadratically and
roughly cubically, respectively. This suggests that the densities of the
simplicial complex models were at most roughly constant, regardless of
the number of landmarks. Landmark covers grew fuzzier and generated more
accurate predictions until the number of landmarks reached around 60,
beyond which point most covers grew crisper while performance increased
more slowly (and in one case decreased). This pattern held for covers
with any fixed multiplicative extension. Naturally, these extensions
produced fuzzier clusters, but they also reduced the overall accuracy of
the predictive model. Independently of these patterns, models fitted to
smaller care units tended to outperform those fitted to larger care
units. Contrary to expectations, unextended maxmin covers were usually
crisper than their lastfirst counterparts and yielded more accurate
predictions, though extensions reduced the crispness of maxmin covers
more dramatically than of lastfirst covers. The same patterns were
observed in the risk discrimination of maxmin versus lastfirst covers,
with maxmin covers yielding the most accurate predictions when
unextended but lastfirst covers retaining more accuracy after extension.

\begin{figure}
\includegraphics[width=.5\textwidth]{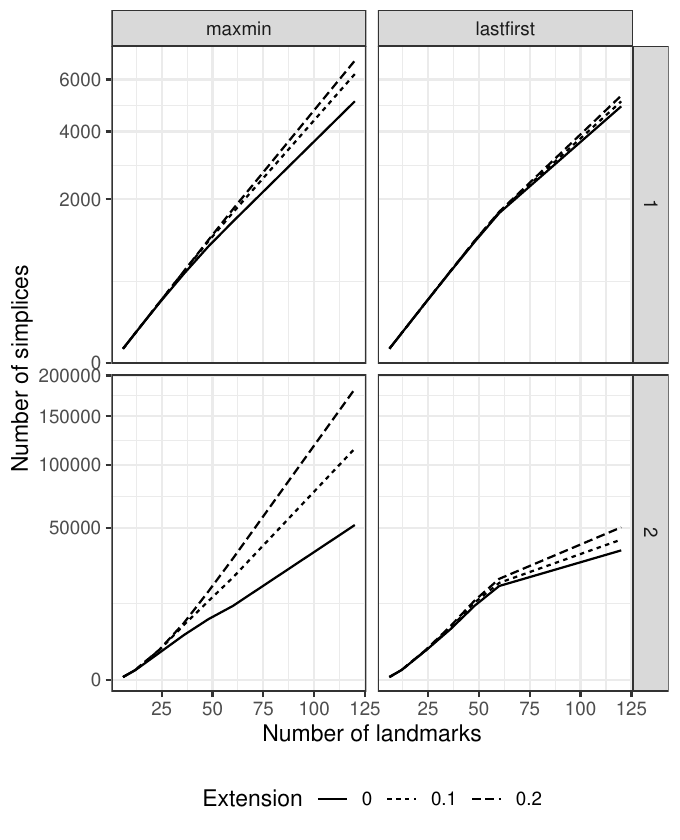}
\includegraphics[width=.5\textwidth]{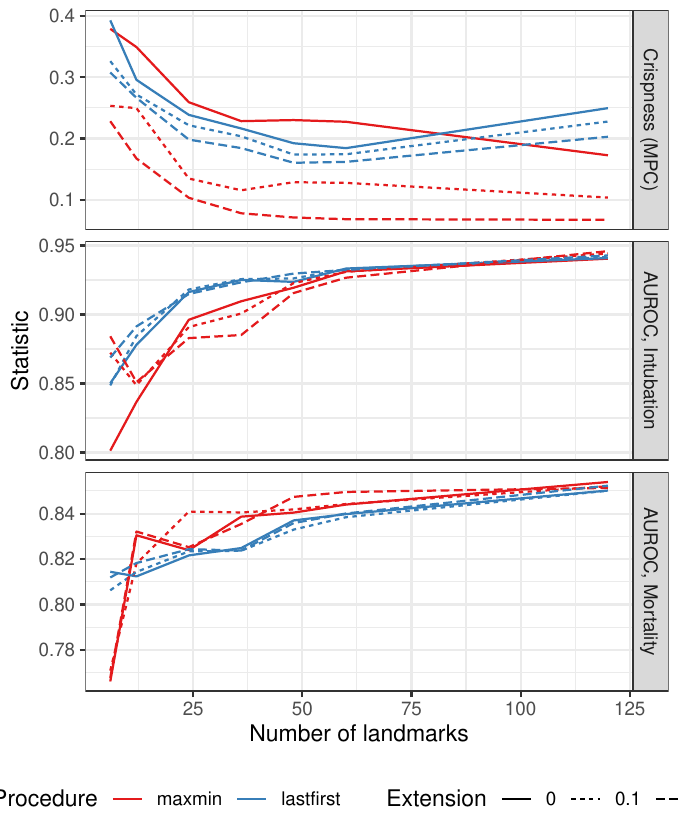}
\caption{
Summary and evaluation statistics versus number of 0-simplices (landmarks) for the covers generated using the maxmin and lastfirst procedures, with three multiplicative extensions in their size.
Left: the sizes of their nerves, as numbers of 1- and 2-simplices, using a square root--transformed vertical scale.
Right: the modified partition coefficient (MPC) and the c-statistics of the risk prediction models based on the cover sets (AUROC).
\label{fig:cover-mx}
}
\end{figure}

Figure\nbs\ref{fig:cover-mx} presents the same evaluations for covers of
the MXDH data. In contrast to the MIMIC experiments, lastfirst-based
nerves of the MXDH data grew sub-polynomially and were significantly
sparser than maxmin-based nerves. Lastfirst covers tended to be crisper,
especially as the number of landmarks and the extension factors
increased. This indicates that the nearest neighborhoods formed a more
parsimonious cover of the data than the centered balls. The predictive
accuracies of the cover set--based models converged with increasing
numbers of landmarks, though for smaller numbers different selection
procedures performed best for different outcomes.

\hypertarget{interpolative-nearest-neighbors-prediction-1}{%
\subsubsection{Interpolative nearest neighbors
prediction}\label{interpolative-nearest-neighbors-prediction-1}}

Both ball and neighborhood covers were used to build INN models of
clinical outcomes. Their relative performance differed on the MIMIC
data, somewhat by unit but most significantly by the choice of metric.
The predictive models performed similarly under the angle distance on
the variables of Lee, Maslove, and Dubin (2015), while lastfirst
outperformed maxmin on the same variables under the Gower distance and
maxmin outperformed lastfirst under the angle distance on the
RT-transformed demographic and diagnosis variables. In every case the
deterministic samplers outperformed random selection. We plot results
using the Gower distance and include analogous plots in the Appendix.

\begin{figure}
\includegraphics[width=\textwidth]{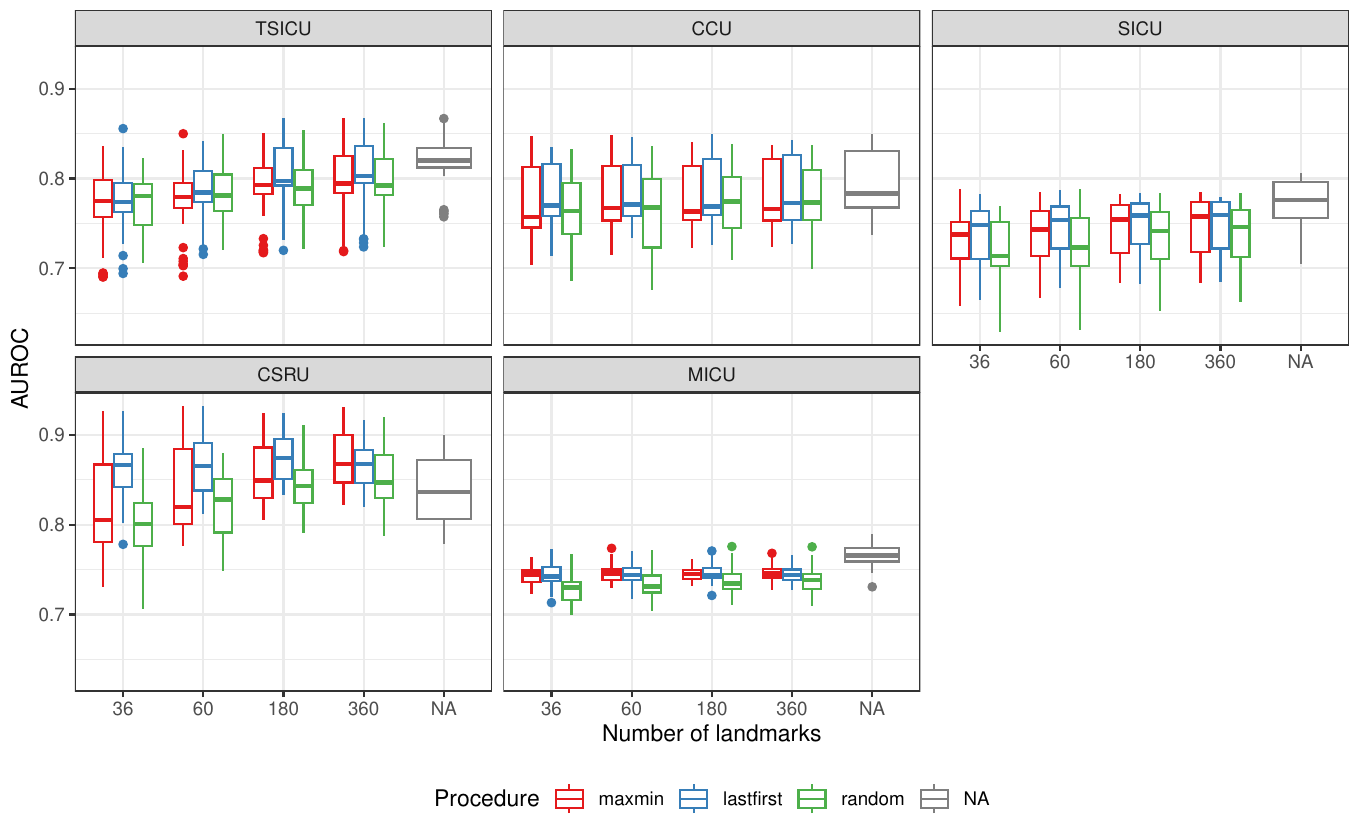}
\caption{
AUROCs of the INN models of mortality in five MIMIC-III care units based on covers constructed using random, maxmin, and lastfirst procedures under Gower distance on domain-informed variables to generate landmarks.
Each boxplot summarizes AUROCs from $6 \times 6 = 36$ models, one for each combination of outer and inner fold.
AUROCs of simple nearest-neighbor predictive models are included for comparison.
\label{fig:knn-mimic-gower}
}
\end{figure}

Boxplots of the AUROCs for each cross-validation step are presented in
Figure\nbs\ref{fig:knn-mimic-gower}. The variation across folds exceeds
the variation between sampling procedures, but the superiority of
lastfirst is visible and consistent throughout, though in some cases one
or both deterministic samplers fail to outperform random selection.
Notably, for several care units, a basic unweighted nearest neighbors
model outperformed the INN models fitted using both samplers. This was
not the case when using the other two distances. INN models on most
units improved performance by incorporating more landmarks, though in no
case models on 360 landmarks clearly outperform those on 180 landmarks.

Over the course of the COVID-19 pandemic, hospitals and other facilities
experienced periods of overburden and resource depletion, and best
practices were continually learned and disseminated. As a result,
outcomes in the MXDH data reflect institutional- as well as
population-level factors. We took advantage of the rapid learning
process in particular by adapting the nested CV approach above to a
temporal CV approach (Major, Jethani, and Aphinyanaphongs 2020): We
partitioned the data by week, beginning with Week\nbs11
(March\nbs11--17) and ending with Week\nbs19 (May\nbs6--9, the last
dates for which data were available). For each week \(i\),
\(11 < i \leq 19\), we trained prediction models on the data from Week
\(i-1\). We then randomly partitioned Week\nbs\(i\) into six roughly
equal parts and optimized and evaluated the models as above. (For this
analysis, we only considered Gaussian weighting.)

\begin{figure}
\includegraphics[width=\textwidth]{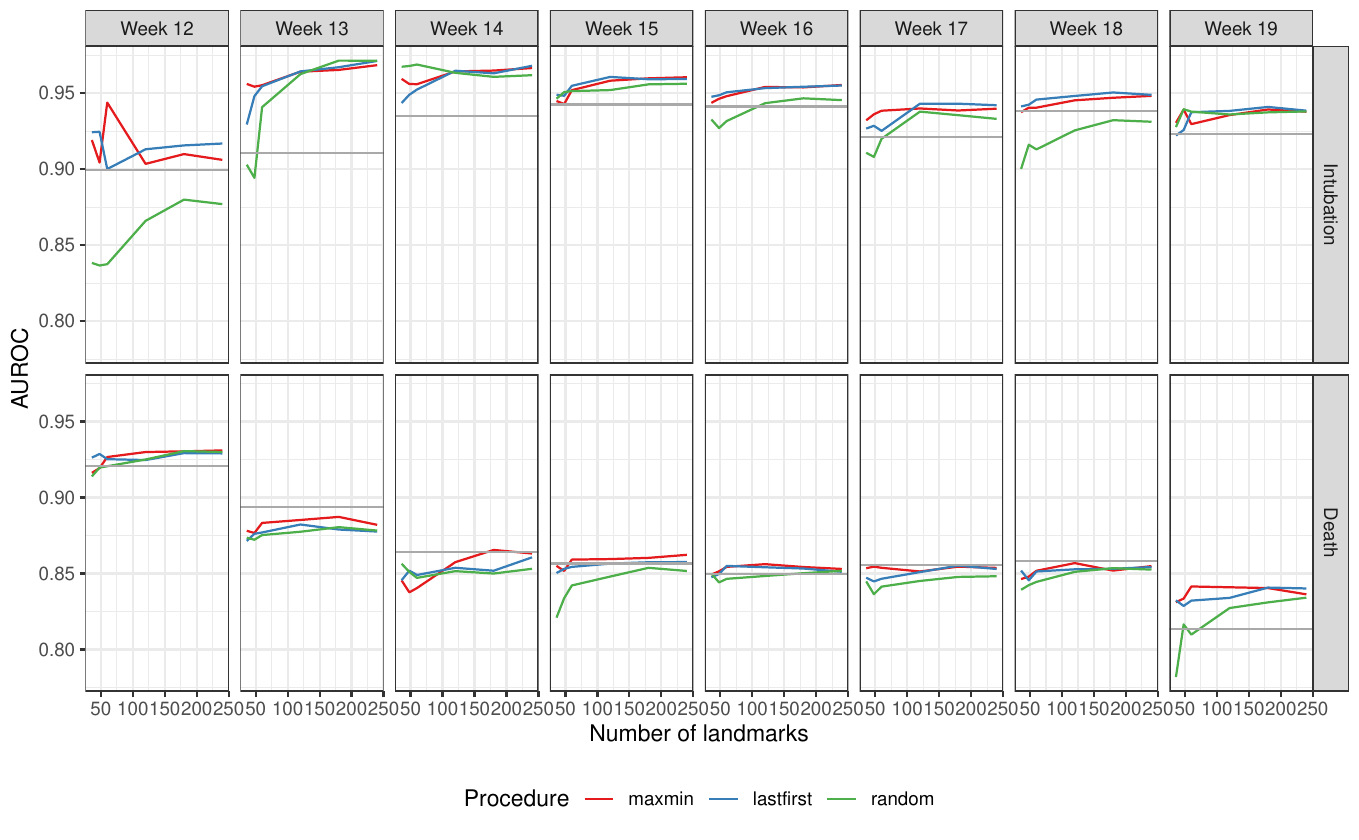}
\caption{
AUROCs of the sliding-window INN models of intubation and mortality in the MXDH data based on covers constructed using random, maxmin, and lastfirst procedures to generate landmarks.
\label{fig:knn-mx}
}
\end{figure}

Line plots of model performance are presented in
Figure\nbs\ref{fig:knn-mx}, with one curve (across numbers of landmarks)
per selection procedure, outcome, and week. Again both landmark
selection procedures yielded stronger results than random selection.
Interestingly, this was more pronounced in later weeks, as the pandemic
progressed, even as overall predictive accuracy declined. Overall,
performance improved slightly as the number of landmarks increased from
50 to 150 but either plateaued or declined from 150 to 250.

\hypertarget{discussion}{%
\section{Discussion}\label{discussion}}

\label{sec:discussion}

In this work, we introduce the lastfirst procedure as a complement to
the widely-used maxmin sampler for the selection of landmark points from
a data set. Lastfirst applies the same logic to the growing of nearest
neighborhoods rather than balls around the sampled points. This results
in a heuristic sample of landmarks and an associated cover with
analogous properties to those of maxmin. Common data limitations require
careful consideration of edge cases, which also apply to maxmin but have
not been worked out before, and our choices of solution can be used with
either procedure. Further, our experiments described in the previous
section demonstrate a number of advantages that the lastfirst sampler
confers over maxmin.

First of all, the lastfirst procedure is more general than maxmin, in
that it can be applied to any set of cases for which directed pairwise
distances are available or can be computed. This relaxes the symmetry
assumption of maxmin and would allow lastfirst to be used when the
relevance of one case to another is not a two-way street. One use case
would be to sample dispersed nodes from a transportation network
involving one-way streets and other asymmetries such as traffic
patterns, which would be modeled as a directed graph with shortest path
distances between nodes. However, it imposes costs to computation, as
our algorithm requires us to either impute or compute and store all
distance ranks from the new landmark at each step. Based on our
experiments comparing implementations of the same type, this increases
runtime and storage only by a constant factor.

Lastfirst outperformed maxmin in settings tailored to the strengths of
the algorithm (and achieved comparable performance in other settings).
On simulated samples from an uneven density function on the circle,
lastfirst-based witness complexes more reliably detected the underlying
0- and 1-dimensional features than maxmin-based ones, and this superior
performance obtained over a wide range of sampler and procedure
parameters. These experiments also demonstrated the value of a simple
additional feature of our implementation, the option to multiplicatively
or additively extend the sizes of cover sets. This generalizable
procedure holds promise for the range of TDA applications like
mapper-type constructions that involve single covers or tunable families
of covers.

Additionally, lastfirst confers some interpretability benefits in the
context of applications. When dealing with certain data types, the
meaning of numerical values of distance may not be clear or intuitive.
For instance, on patient medical data obtained from electronic health
records, custom distance metrics are often used to measure the
\emph{relative} clinical relevance between patients, and it may not be
obvious what an \emph{absolute} distance of, say, \(.35\) between
patient A and patient B might mean. In these types of settings, it can
therefore be difficult to select parameters such as a ball radius to use
in maxmin or other algorithms, and such choices are often made
arbitrarily. This is especially true when minimal literature references
are available, as is the case for many non-mathematical applications of
TDA. However, in most applications, the notion of neighborhoods remains
intuitive: Even when a ball of radius \(.35\) around patient A is hard
to conceptualize, a neighborhood of 200 similar patients still has a
clear meaning. This increased interpretability makes parameter selection
easier and also makes the algorithm more accessible to researchers
outside mathematics.

We ran several experiments that used landmarks to obtain well-separated
clusters of patients with common risk profiles and to more efficiently
generate nearest neighbor predictions. Because we designed lastfirst to
produce cover sets of equal size despite variation in the density or
multiplicity of the data, we expected it to outperform maxmin with
respect to the crispness of clusters and to the accuracy of predictions.
In particular, we expected that the optimal neighborhood size for
outcome prediction would be roughly equal across our data; as a result,
by assigning each landmark case an equally-sized cohort of similar
cases, we expected predictions based on these cohorts to outperform
those based on cohorts using a fixed similarity threshold.

Contrary to expectations, maxmin produced crisper clusterings on
average; though, when the clusters were extended by a fixed proportion,
those of lastfirst better preserved these qualities. Which localized
cohorts were more predictive of clinical outcomes---those obtained using
maxmin-based ball covers or lastfirst-based neighborhood
covers---varied. When predicting mortality on MIMIC-III, the answer
depended largely on the care unit but also, importantly, on the
underlying distance measure. In the case of MXDH, neither landmark
selection procedure produced consistently more accurate predictions.

Under certain conditions, then, the use of personalized cohorts to
improve predictive modeling, as employed by Lee, Maslove, and Dubin
(2015), may be strengthened by optimizing a fixed similarity threshold
rather than a fixed cohort size. It is worth noting that Park, Kim, and
Chun (2006), to our knowledge the only other investigators who have
compared predictive models based on cohorts bounded by a radius versus a
cardinality, reached a similar conclusion. The relative performance of
lastfirst to maxmin was greater on most tasks involving the MXDH data,
which had more categorical variables, more indistinguishable records,
and fewer variables overall---that is, data for which the limitations of
maxmin we described at the outset are more acute. Yet it was least at
predicting mortality in MIMIC-III when similarity was calculated
exclusively from binary variables (RT-transformed data).

On the whole, lastfirst covers are competitive with maxmin covers for
basic analysis tasks. In the absence of clear determinants of
superiority, we suggest that lastfirst should be considered for more
advanced tasks, for example witness complexes or mapper-like
constructions, in cases where consistency in the sizes of cover sets or
density-based representativeness is itself advantageous, e.g.~when
generating a sample for downstream modeling purposes.

One way to reconcile our results is in terms of a balance between
relevance and power, with fixed-radius balls (respectively,
fixed-cardinality neighborhoods) providing training cohorts of roughly
equal relevance (statistical power) to all test cases. With sufficiently
rich data, relevance can be more precisely measured and becomes more
important to cohort definition, as with MIMIC. When variables are fewer,
as with MXDH, relevance is more difficult to measure, so that larger
samples can improve performance even at the expense of such a measure.

\hypertarget{appendix}{%
\section{Appendix}\label{appendix}}

\hypertarget{relative-ranks}{%
\subsection{Relative ranks}\label{relative-ranks}}

This section develops a more general lastfirst procedure and makes
rigorous some ideas in the main text.

Relative ranks are a much more general notion of metric that encompasess
ranks in nearest neighborhoods.

\begin{definition}[relative rank]
    A \emph{relative rank} on $X$ is a binary relation $q: X \times X \to \bbR_{\geq 0}$ subject only to the following inequality:
    \begin{equation}\label{eqn:relative-rank}
        \forall\, x,y \in X : d(x,x) \leq d(x,y)
    \end{equation}
\end{definition}

Relative ranks can be used as the basis for a much more general maxmin
procedure, taking care in particular to account for asymmetry.

Given a relative rank \(q\) on \(X\), write \begin{align*}
    q(Y,Z) &= \min_{y\in Y,z\in Z}{q(y,z)} & Q(Y,Z) &= \max_{y\in Y,z\in Z}{q(y,z)} \\
    q(x,Y) &= q(\{x\},Y)                   & Q(x,Y) &= Q(\{x\},Y) \\
    q(Y,x) &= q(Y,\{x\})                   & Q(x,Y) &= Q(Y,\{x\}) \\
\end{align*}

A pseudometric \(d_X\) induces a relative rank that takes values in
\(\bbN\) given by the ordinal of one point's distance from another:

\begin{definition}[out-rank and in-rank]
    For $x,y\in X$ with pseudometric $d$, define the \emph{out-rank} $q_{X,d} : X \times X \longrightarrow \bbN$ as follows:
    \begin{equation}\label{eqn:out-rank}
        q_{X,d}(x,y)=\abs{\{z\in X \where d(x,z) < d(x,y)\}}
    \end{equation}
    and the \emph{in-rank} $q_{X,d}^\top(x,y) = q_{X,d}(y,x)$.
\end{definition}

As with \(d\), we allow ourselves to write \(q=q_X=q_d=q_{X,d}\) when
clear from context. Note that \(q_{X,d}\) is a relative rank with
\(q(x,x)=0\), \(q(x,y) < N\), and
\(\forall\, x,y \in X : q(x,x) \leq q(x,y)\).

A \emph{quasi-pseudometric} that violates symmetry, or a
\emph{semi-(quasi-)pseudometric} that satisfies a weaker triangle
inequality, induces a relative rank in the same way, since in each case
the axiom \(d(x,x) = 0\) provides that
\(q_d(x,x) = \abs{\{z\in X \where d(x,z) < d(x,x)\}} = 0 \leq q_d(x,y)\).
More interestingly, if this axiom is dropped to define a \emph{partial
pseudometric} (Matthews 1994), then the axiom \(d(x,x) \leq d(x,y)\)
(named ``small self-distances'\,' by Bukatin et al. (2009)) likewise
provides Equation\nbs\ref{eqn:relative-rank}.

It is also worth noting that the out-rank construction is idempotent.
This follows from the observation that \(q_d(x,z) < q_d(x,y)\) means
that \(\exists\, w : d(x,z) \leq d(x,w) < d(x,y)\) and can be
reformulated simply as \(d(x,z) < d(x,y)\). When \(d\) is itself a
relative rank induced from a generalized metric, then both \(d\) and
\(q_d\) are completely determined by these inequalities
(Equation\nbs\ref{eqn:out-rank}).

\begin{example}\label{ex:relative-rank}
    Recall $X=\{a,b,c,d\}$ from Example\nbs\ref{ex:rank-sequence-order}.
    Lack of symmetry of $q$ is shown by points $b$ and $c$ :
    \begin{align*}
        q(a,c) &= \abs{\{x \in X \where \abs{x-a} < \abs{c-a} = 2\}} &
        q(c,a) &= \abs{\{x \in X \where \abs{x-c} < \abs{a-c} = 2\}} \\
               &= \abs{\{a, b\}} &&= \abs{\{b, c, d\}} \\
               &= 2 &&= 3
    \end{align*}

    Observe in particular that $\max_{x\in X}{q(a,x)}=2<\abs{X}-1$; $q(x,\,\cdot\,)$ will not max out at $N-1$ when the most distant points from $x$ have multiplicity.

    However, $q(x,x) = 0$ whether $x$ has multiplicity or not:
    \begin{align*}
        q(a,a) &= \abs{\{x \in X \where \abs{x-a} < \abs{a-a} = 0\}} &
        q(c,c) &= \abs{\{x \in X \where \abs{x-c} < \abs{c-c} = 0\}} \\
               &= \abs{\varnothing} &&= \abs{\varnothing} \\
               &= 0 &&= 0
    \end{align*}
\end{example}

We also term the unary rankings \(q(x,\,\cdot\,)\) and
\(q(\,\cdot\,,x)\) the \emph{out- (from $x$)} and
\emph{in- (to $x$) rankings} of \(X\), respectively. These can be used
to define \emph{out-} and \emph{in-neighborhoods} of \(x\).\footnote{The
  terminology and notation are adapted from graph theory. These
  definitions are the same as those for a complete directed graph on
  \(X\) with directed arcs \(x\to y\) weighted by \(q(x,y)\).}

A relative rank \(q\) can be used to define \(k\)-neighborhoods in
greater generality:

\begin{definition}[$k$-neighborhoods using relative rank]
    For $x \in X$, define the \emph{$k$-out-neighborhoods} $N^+_k$ and \emph{$k$-in-neighborhoods} $N^-_k$ of $x$ as the sets
    \begin{align*}
        & N^+_k(x)=\{y\in X \where q(x,y)\leq k\} \\
        & N^-_k(x)=\{y\in X \where q(y,x)\leq k\}
    \end{align*}
    Given a subset $Y \subseteq X$, we also define
    \begin{align*}
        & N^+_k(x,Y)=\{y\in Y \where q_X(x,y)\leq k\} \\
        & N^-_k(x,Y)=\{y\in Y \where q_X(y,x)\leq k\}
    \end{align*}
\end{definition}

Relative ranks are not as straightforward to compare among subsets of
points. For example, for \(y\neq x\), \(q(x,y)\) takes integer values
between \(\abs{\supp{\{x\}}}\) and \(N-1\). However, they do provide us
with a definition of the lastfirst procedure that straightforwardly
adapts Definition\nbs\ref{def:maxmin}.

\begin{corollary}[lastfirst using relative rank]
    Given $Y\subset X$ and a pseudometric $d$ on $X$ with relative rank $q$,
    \begin{align*}
        \lf(Y;X) &= \maxmin(Y;X,q_d) \\
        \lf(X,d) &= \maxmin(X,q_d)
    \end{align*}
\end{corollary}

\begin{proof}
This follows directly from the observations
\begin{align*}
    \maxmin(Y;X,q) &= \{x\in X\wo \cl{Y} \where N_\bullet^-(x,Y) = \max_{y\in X\wo \cl{Y}}{N_\bullet^-(y,Y)}\} \\
    \maxmin(X,q) &= \{x\in X \where N_\bullet^-(x,X \wo \cl{\{x\}}) = \max_{y\in X}{N_\bullet^-(y,X \wo \cl{\{y\}}}\}
\end{align*}
obtained by adapting Proposition\nbs\ref{prop:maxmin} to the relative rank.
\end{proof}

\hypertarget{selection-procedures}{%
\subsubsection{Selection procedures}\label{selection-procedures}}

The choice of \(\ell_i \in \maxmin(L)\) is trivial when \(X\) is in
locally general position, but this study is specifically interested in
cases with a high frequency of violations of this property, and indeed
with such violations of Hausdorffness that large numbers of points in
\(X\) may be co-located. When \(X\) is not in locally general position,
the choice of selection from among the maxmin set is consequential. For
convenience, let \(\maxmin(L;X)\) take the value \(\varnothing\) if
\(\cl{L}=X\) and the value \(X\) if \(L=\varnothing\). Then write
\(\graph{\maxmin, X} \subset \order{X} \times \power{X}\) for the graph
of the unary function \(\maxmin(\,\cdot\,;X): \order{X} \to \power{X}\),
so that \((L,\Gamma) \in \graph{\maxmin, X}\) if and only if
\(\maxmin(L;X) = \Gamma\). Then a selection procedure is a function
\(\sigma: \graph{\maxmin, X} \to X\) subject to
\(\sigma((L,\Gamma)) \in \Gamma = \maxmin(L;X)\). Importantly,
\(\sigma\) may depend not only on the maxmin set \(\Gamma\) but also on
the ordered sequence \(L\).

We assume the following choice of \(\sigma\): Take
\(d_L = \max_{y \in X \wo \cl{L}}{d(y,L)}\). For each
\(y \in \maxmin(L;X)\), choose \(\ell_y \in L\) for which
\(d(y,\ell_y) = d_L\). Then take
\(\maxmin^{(1)}(L;X) = \{y \in \maxmin(L;X) \where d(y,L \wo \{\ell_y\}) = \max_{z \in \maxmin(L;X)}{d(z,L \wo \{\ell_z\})}\}\).
While \(\abs{\maxmin^{(j)}(L;X)} > 1\), continue in this way until
either a singleton is reached or \(j=\abs{L}=i\). If the latter, then
the choice \(\sigma\) is arbitrary among the remaining \(y\).

Suppose \(\ell_i\) is selected so that the minimum \(k\) required for
\(\ell_i \in \bigcup_{j=0}^{i-1}{N_k(\ell_j)}\) is maximized. This is
equivalent to maximizing the minimum out-rank \(q(\ell_j,\ell_i)\) of
\(\ell_i\) from any \(\ell_j\). Switching perspective from out- to in-
and reversing the roles of \(L\) and \(\ell_i\), we want
\(N^-_k(\ell_i,L)=0\) for the latest (largest) \(k\) possible, say
\(k^-_0\). When indistinguishable points abound, this may still not
uniquely determine \(\ell_i\), so we may extend the principle: Among
those \(\ell\in X\wo \cl{L}\) for which \(N^-_{k^-_0}(\ell_i,L)=0\),
choose \(\ell_i\) for which \(N^-_{k}(\ell_i,L) \leq 1\) for the latest
\(k\) possible, say \(k^-_1 \geq k^-_0\). (It is possible that
\(k^-_1 = k^-_0\), in which case no \(N^-_k(\ell_i,L) = 1\).) Continue
this process until only one candidate \(\ell\) remains (up to
multiplicity), or until \(N^-_{k}(\ell,L)=\abs{L}\), in which case all
remaining candidates may be considered equivalent.

\hypertarget{algorithms}{%
\subsubsection{Algorithms}\label{algorithms}}

Algorithm\nbs\ref{alg:lastfirst-landmarks} calculates a lastfirst set
from a seed point, subject to parameters analogous to \(n\) and \(\eps\)
in Algorithm\nbs\ref{alg:maxmin}. The algorithm is tailored to the
vectorized arithmetic of \pkg{R}, and Lemma\nbs\ref{lem:revlex-lex}
provides a shortcut between \(Q^-\) and the more compact way that the
relative rank data are stored.

\begin{lemma}\label{lem:revlex-lex}
For $L = \{ \ell_0, \ldots, \ell_n \} \subset X$, write $S(x,L) = ( q(\ell_{\pi^{-1}(1)},x) \leq \cdots \leq q(\ell_{\pi^{-1}(n)},x) )$, where $\pi$ is any suitable permutation on $[n]$.
Then $Q^-(x,L) < Q^-(y,L) \Leftrightarrow S(x,L) < S(y,L)$.
\end{lemma}

\begin{proof}
Write $Q(x) = Q^-(x,L)$ and $Q(y) = Q^-(y,L)$ and suppose that $Q(x) < Q(y)$.
This means that $Q_i(x) > Q_i(y)$ for some index $i \in [N]$ while $Q_j(x) = Q_j(y)$ for all $j < i$.
There are then, for each $j < i$, equal numbers of $\ell \in L$ for which $q(\ell,x) = j$ and for which $q(\ell,y) = j$; while there are more $\ell \in L$ for which $q(\ell,x) = i$ than for which $q(\ell,y) = i$.
When the sets $\{ q(\ell,x) \}_{\ell in L}$ and $\{ q(\ell,y) \}_{\ell in L}$ are arranged in order to get $S(x,L)$ and $S(y,L)$, therefore, the leftmost position at which they differ is $Q_1(y) + \cdots + Q_i(Y) + 1$, at which $q(\ell,x) = i$ while $q(\ell,y) \geq i + 1$.
Thus $S(x,L) <_{\operatorname{lex}} S(y,L)$.

The reverse implication is similarly straightforward.
\end{proof}

\begin{algorithm}
\caption{Calculate the lastfirst landmark sequence from a seed point.}
\label{alg:lastfirst-landmarks}
\begin{algorithmic}[1]
\REQUIRE finite pseudometric space $(X,d)$
\REQUIRE seed point $\ell_0 \in X$
\REQUIRE number of landmarks $n \in \bbN$ or cover set cardinality $k \in \bbN$
\REQUIRE selection procedure $\sigma$
\STATE if $n$ is not given, set $n \leftarrow 0$
\label{line:n}
\STATE if $k$ is not given, set $k \leftarrow \infty$
\label{line:k}
\STATE $L \leftarrow \varnothing$ initial landmark set
\STATE $F \leftarrow \{ \ell_0 \}$ initial lastfirst set
\STATE $R \in \bbN^{N \times 0}$, a $0$-dimensional $\bbN$-valued matrix
\FOR{$i$ from $0$ to $\uniq{X} - 1$}
    \STATE $\ell_i \leftarrow \sigma(F)$
    \STATE $L \leftarrow L \cup \{\ell_i\}$
    \STATE $D_i \leftarrow (d_{i1},\ldots,d_{iN}) \in {\bbR_{\geq 0}}^N$, where $d_{ir} = d(\ell_i, x_r)$
    \STATE $Q_i \leftarrow \verb|rank|(D_i) \in {\bbN_{\geq 0}}^N$ (so that $Q = (q(\ell_i, x_1),\ldots,q(\ell_i, x_N))$)
    \label{line:rank}
    \STATE $R \leftarrow [R, Q_i] \in \bbN^{N \times (i+1)}$
    \STATE $k_{\min} \leftarrow \max_{r=1}^{N}{ \min_{j=1,i+1}{ R_{r,j} } }$ (minimum $k$ for which neighborhoods centered at $L$ cover $X$)
    \label{line:kmin}
    \IF{$D(L, X \wo \cl{L}) = 0$}
        \STATE \textbf{break}
        \label{line:nonempty}
    \ENDIF
    \IF{$i \geq n$ and $k_{\min} \leq k$}
        \STATE \textbf{break}
        \label{line:check}
    \ENDIF
    \STATE $R \leftarrow [ \verb|sort|({R_{1,\bullet}})^\top \cdots \verb|sort|({R_{N,\bullet}})^\top ]^\top \in \bbN^{N \times (i+1)}$
    \label{line:sort}
    \STATE $F \leftarrow X \wo \cl{L}$
    \FOR{$j$ from $1$ to $i$}
    \label{line:maximize}
        \STATE $F \leftarrow \{x_r \in F \where R_{rj} = \max_{r'}{R_{r'j}}\}$
        \label{line:lastfirst}
        \IF{$\abs{F} = 1$}
            \STATE \textbf{break}
        \ENDIF
    \ENDFOR
\ENDFOR
\RETURN $L$
\RETURN lastfirst landmark set $L$ with at least $n$ cover sets of cardinality at most $k$
\end{algorithmic}
\end{algorithm}

\begin{proposition}
Algorithm\nbs\ref{alg:lastfirst-landmarks} returns a lastfirst landmark set.
If $n \leq \uniq{X}$ is given as input and $k$ is not, then $\abs{L} = n$.
If $n$ and $k$ are both given, then $\abs{L} \geq n$.
Otherwise, $L$ is minimal in the sense that no proper prefix of $L$ gives a cover of $X$ by $k$-nearest neighborhoods.
\end{proposition}

\begin{proof}
Let $(X,d)$ be a finite metric space and $\ell_0 \in X$ be a seed point, as required by Algorithm\nbs\ref{alg:lastfirst-landmarks}.
Note that, for the algorithm to terminate its loop and subsequently return $L$, either there must be no points in $X \wo \cl{L}$ distinguishable from $L$ (line\nbs\ref{line:nonempty}), or both of two exit conditions must hold (line\nbs\ref{line:check}):
  (1) $k_{\min} \leq k$ and
  (2) $\abs{L} \geq n$.

Because the seed point is arbitrary, for the main result it is enough to show that, at each step $i$, $F = \lf(\{ \ell_0, \ldots, \ell_{i-1} \})$.
When $F$ is calculated on line\nbs\ref{line:lastfirst}, the rows $R_r$ of $R$ contain the in-ranks $q(\ell_i, x_r)$ of $x_r$ in increasing order.
Because $D(L, X \wo \cl{L}) > 0$ (line\nbs\ref{line:nonempty}), $F$ is nonempty.
By Lemma\nbs\ref{lem:revlex-lex}, then, $Q^-(x_r, L)$ is maximized (in revlex) when $R_r$ is maximized in lex, and this is exactly what the loop that begins on line\nbs\ref{line:maximize} does.

Suppose first that $n \leq \uniq{X}$ is given.
The loop will only break on line\nbs\ref{line:nonempty} if $D(L, X \wo \cl{L}) = 0$, which is only possible if $\abs{L} = \uniq{X} \geq n$.
The loop will only break on line\nbs\ref{line:check} if both $\abs{L} = i \geq n$ and $k_{\min} \leq k$.
Since these are the only two possible breaks, $\abs{L} \geq n$ is a necessary condition.
\footnote{Note that $\abs{L} > n$ if and only if $k_{\min} \leq k$ is not satisfied when $\abs{L} = n$, meaning that $X$ would not be covered by $k$-neighborhoods around $n$ landmark points, so that more landmarks must be chosen to guarantee the algorithm produces a valid $k$-neighborhood cover.}

If $k$ is not given, then $k$ is set to $\infty$ on line\nbs\ref{line:k}, which means that (1) holds throughout the loop.
Then the algorithm terminates as least as soon as (2) is satisfied, when $\abs{L} = n$, and as discussed above it cannot terminate any sooner.

Now suppose $n$ is not given.
Then $k$ must be given, and $n$ is set to $0$ (line\nbs\ref{line:n}).
This means that (2) always holds, so the algorithm terminates as soon as (1) is satisfied, i.e.\ when $k \geq k_{\min}$ with $k_{\min}$ as defined on line\nbs\ref{line:kmin}.
We claim that
\[ k_{\min} = \max_{x \in X}{ \min_{\ell \in L}{ q(\ell, x) } } \]
which means that every point in $x$ is within a $k$-neighborhood of some existing landmark $\ell \in L$, i.e.\ that the $k$-neighborhoods at $L$ constitute a cover of $X$.
For this to have not been true for $L$ at previous iterations, there must have been $x \in X$ with $q(\ell, x) > k$ for all $\ell \in L$, meaning that the $k$-neighborhoods at $L$ did not cover $X$.

To prove the claim, note that the maximum is taken over all rows of $R$, which at no point in the algorithm are permuted---that is, the entries of row $R_{r,\bullet}$ at each iteration consist of $q(\ell_0,x_r),\ldots,q(\ell_i,x_r)$ in some order.
Therefore, $\min_{j=1}^{i}{R_{r,j}} = \min_{\ell \in L}{ q(\ell, x_r) }$.
Because columns $1$ through $i-1$ of $R$ were sorted in the previous iteration (line\nbs\ref{line:sort}), this minimum only needs to be taken over $j=1,i$, which gives the formula on line\nbs\ref{line:kmin}.
\end{proof}

\hypertarget{tie-handling}{%
\subsubsection{Tie handling}\label{tie-handling}}

We might have defined two relative ranks
\(\check{q}, \hat{q} : X \times X \longrightarrow \bbN\)
(``\(q\)-check'' and ``\(q\)-hat'') as follows: \begin{align*}
& \check{q}(x,y)=\abs{\{z\in X \where d(x,z)<d(x,y)\}} \\
& \hat{q}(x,y)=\abs{\{z\in X \where d(x,z)\leq d(x,y)\}} - 1
\end{align*} In this notation, \(\check{q}=q\), while \(\hat{q}(x,y)\)
is the cardinality of the smallest ball centered at \(x\) that contains
\(y\). Then
\(\check{N}^\pm_1(x) \subseteq \{x\} \subseteq \hat{N}^\pm_1(x)\), and
\(\hat{q}(x,x)>0\) when \(x\) has multiplicity. The two relative ranks
derive from two tie-handling schemes for calculating rankings of lists
with duplicates. For example, if \(a<b=c<d\) are the distances from
\(x\) to \(y_1,y_2,y_3,y_4\), respectively, then
\((\check{q}(x,y_1),\check{q}(x,y_2),\check{q}(x,y_3),\check{q}(x,y_4))=(0,1,1,3)\)
and
\((\hat{q}(x,y_1),\hat{q}(x,y_2),\hat{q}(x,y_3),\hat{q}(x,y_4))=(0,2,2,3)\).
Indeed, any tie-handling rule could be used, and the choice becomes more
consequential with greater multiplicity in the data.

Conceptually, the lastfirst procedure based on \(\hat{q}\) would produce
landmark sets that yield neighborhood covers with smaller, rather than
larger, neighborhoods in regions of high multiplicity. While we do not
use these ideas in this study, they may be suitable in some settings or
for some purposes, for example when high multiplicity indicates a
failure to discriminate between important categories. It is also
possible that \(\check{q}\)- and \(\hat{q}\)-based covers could be used
to produce interveaving sequences of nerves useful for stability
analysis.

\begin{example}\label{ex:relative-rank-max}
Consider the same $X$ as in Example\nbs\ref{ex:relative-rank}.

The relative rank $\hat{q}$ is also asymmetric:
\begin{align*}
    \hat{q}(b,c) &= \abs{\{x \in X \where \abs{x-b} \leq \abs{c-b} = 2\}} &
    \hat{q}(c,b) &= \abs{\{x \in X \where \abs{x-c} \leq \abs{b-c} = 2\}} \\
        &= \abs{\{a, b, c, d\}} &&= \abs{\{b, c, d\}} \\
        &= 4 &&= 3
\end{align*}

Observe that $\hat{q}(x,x) = 1$ only for distinguishable points $x \in X$:
\begin{align*}
    \hat{q}(a,a) &= \abs{\{x \in X \where \abs{x-a} \leq \abs{a-a} = 0\}} &
    \hat{q}(c,c) &= \abs{\{x \in X \where \abs{x-c} \leq \abs{c-c} = 0\}} \\
       &= \abs{\{a\}} &&= \abs{\{c, d\}} \\
       &= 1 &&= 2
\end{align*}

Finally, observe that $\hat{q}(x,\,\cdot\,)$ always maxes out at $\abs{X}$: $\hat{q}(a,c) = \hat{q}(b,c) = \hat{q}(c,a) = \hat{q}(d,a) = \abs{X}$.

Continuing on as in Example\nbs\ref{ex:rank-neighborhoods}, we can compute $\hat{N}_2^+$ and $\hat{N}_2^-$ for $b$ and $c$:
\begin{align*}
    \hat{N}_2^+ (b) &= \{x \in X \where \hat{q}(b,x) \leq 2\} &
    \hat{N}_2^+ (c) &= \{x \in X \where \hat{q}(c,x) \leq 2\} \\
        &= \{a, b\} &&= \{c, d\} \\
        \\
    \hat{N}_2^- (b) &= \{x \in X \where \hat{q}(x,b) \leq 2\} &
    \hat{N}_2^- (c) &= \{x \in X \where \hat{q}(x,c) \leq 2\} \\
        &= \{a, b\} &&= \{c, d\}
\end{align*}

Similarly, we can compute the other $\hat{N}_k^+$ and $\hat{N}_k^-$ for $b$ and $c$:
\begin{align*}
    \hat{Q}^+ (b) &= (\abs{\hat{N}^+_1(b)}, \abs{\hat{N}^+_2(b)}, \abs{\hat{N}^+_3(b)}, \abs{\hat{N}^+_4(b)}) &
    \hat{N}_\bullet^+ (c) &= (\abs{\hat{N}^+_1(c)}, \abs{\hat{N}^+_2(c)}, \abs{\hat{N}^+_3(c)}, \abs{\hat{N}^+_4(c)}) \\
        &= (\abs{\{b\}}, \abs{\{a, b\}}, \abs{\{a, b\}}, \abs{\{a, b, c, d\}}) &
        &= (\abs{\varnothing}, \abs{\{c, d\}}, \abs{\{b, c, d\}}, \abs{\{a, b, c, d\}}) \\
        &= (1, 2, 2, 4) &
        &= (0, 2, 3, 4) \\
        \\
    \hat{N}_\bullet^- (b) &= (\abs{\hat{N}^-_1(b)}, \abs{\hat{N}^-_2(b)}, \abs{\hat{N}^-_3(b)}, \abs{\hat{N}^-_4(b)}) &
    \hat{N}_\bullet^- (c) &= (\abs{\hat{N}^-_1(c)}, \abs{\hat{N}^-_2(c)}, \abs{\hat{N}^-_3(c)}, \abs{\hat{N}^-_4(c)}) \\
        &= (\abs{\{b\}}, \abs{\{a, b\}}, \abs{\{a, b, c, d\}}, \abs{\{a, b, c, d\}}) &
        &= (\abs{\varnothing}, \abs{\{c, d\}}, \abs{\{c, d\}}, \abs{\{a, b, c, d\}}) \\
        &= (1, 2, 4, 4) &&= (0, 2, 2, 4)
\end{align*}
\end{example}

\hypertarget{homology-recovery}{%
\subsection{Homology recovery}\label{homology-recovery}}

We compared the suitability of three landmarking procedures (uniformly
random, maxmin, lastfirst) on datasets with varying density and
duplication patterns by extending an example of de Silva and Carlsson
(2004). Each expriment proceeded as follows: We sampled \(n=540\) points
from the sphere \(\Sph^2\subset\bbR^3\) and in different experiments
selected \(k=12,36,60\) landmark points. We then used the landmarks to
compute PH and computed the \emph{relative dominance}
\((R_1 - R_0) / K_0\) and \emph{absolute dominance}
\((R_1 - R_0) / K_1\) of the last interval over which all Betti numbers
agreed with those of \(\Sph^2\). These statistics provide an indication
of how successfully PH recovered the homology of the manifold from which
the points were sampled.

The points \(x=(r,\theta,\phi)\) were sampled using four procedures:
uniform sampling, skewed sampling, uniform sampling with skewed
boosting, and skewed sampling with skewed boosting. The first procedure
was used by de Silva and Carlsson (2004) and here serves as a baseline.
For a sample \(S\) (with multiplicities) generated from each of the
other three procedures, the expected density
\(\lim_{\eps\to 0}\lim_{n\to\infty}\frac{1}{n}\abs{\{x=(r,\theta,\phi)\in S \where \alpha-\eps<\phi<\alpha+\eps\}}\)
of points near a given latitude \(\alpha\in[0,\pi]\) is proportional to
the quartic function \(p:[0,1]\to[0,1]\) defined by
\(p(x)=(\frac{\phi}{\pi})^4\). Skewed sampling is performed via
rejection sampling: Points \(x_i=(r_i,\theta_i,\phi_i)\) are sampled
uniformly and rejected at random if a uniform random variable
\(t_i\in[0,1]\) satisfies \((\frac{\phi_i}{\pi})^\alpha<t_i\) until
\(n\) points have been kept (Diaconis, Holmes, and Shahshahani 2013).
Skewed boosting is performed by first obtaining a (uniform or skewed)
sample \(T\) of size a fraction \(\frac{n}{6}\) of the total, then
sampling \(n\) points (with replacement) from \(T\) using the
probability mass function satisfying
\(P(x_i)\propto(\frac{\phi_i}{\pi})^\beta\). When performed separately,
skewed sampling and skewed boosting use \(\alpha=\beta=4\); when
performed in sequence, they use \(\alpha=\beta=2\).

The landmark points were selected in three ways: uniform random
selection (without replacement), the maxmin procedure, and the lastfirst
procedure. We computed PH in \pkg{Python GUDHI}, using three
implementations: Vietoris--Rips (VR) filtrations on the landmarks, alpha
complexes on the landmarks, and witness complexes on the landmarks with
the full sample as witnesses (Maria et al. 2021; Rouvreau 2021;
Kachanovich 2021).

The skewed data sets are dense at the south pole and sparse at the north
pole. We expect lastfirst to be more sensitive to this variation and
place more landmarks toward the south. As measured by dominance,
therefore, we hypothesized that lastfirst would be competitive with
maxmin when samples are uniform and inferior to maxmin when samples are
skewed. Put differently, we expected lastfirst to better reject the
homology of \(\Sph^2\), i.e.~to detect the statistical void at the north
pole.

\begin{figure}
\includegraphics[width=\textwidth]{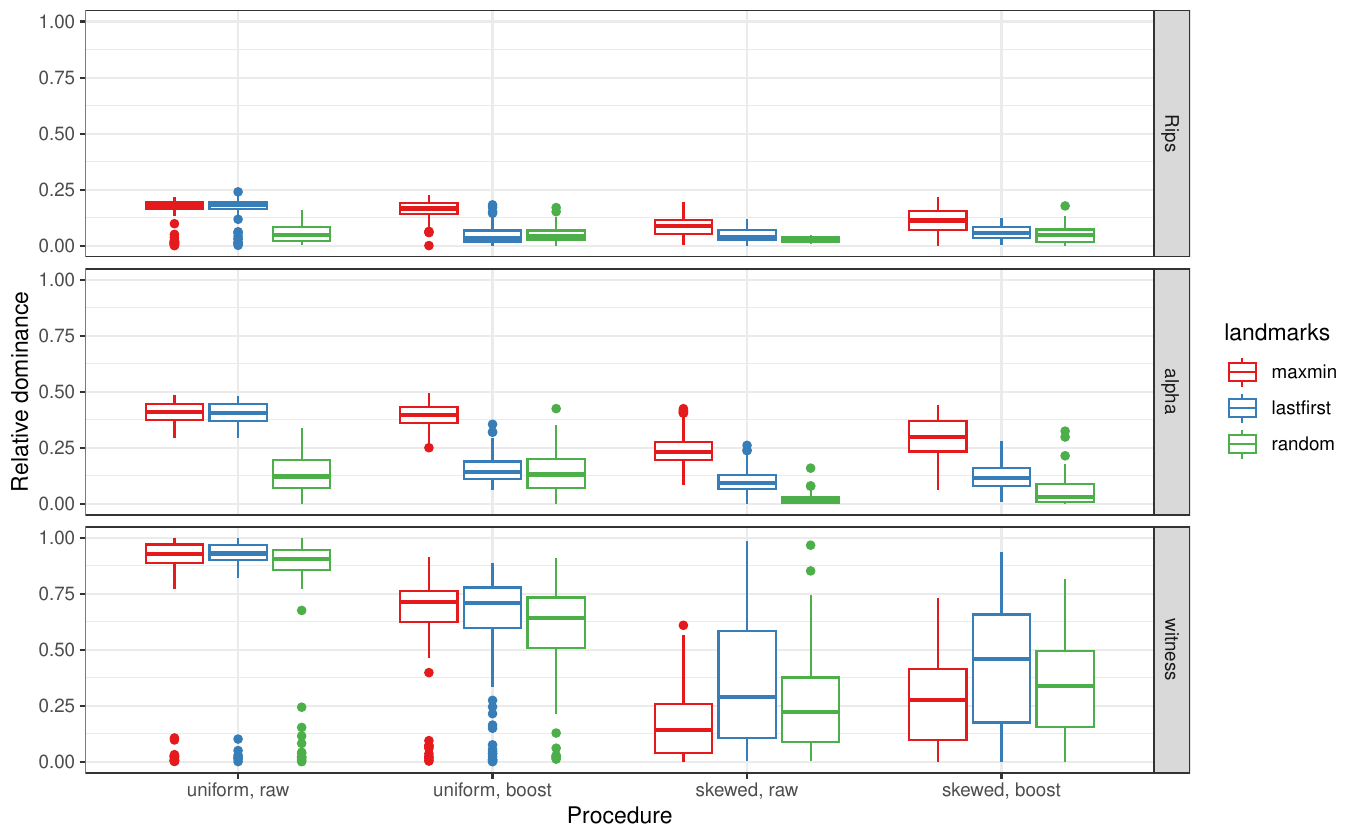}
\caption{
Relative dominance of the spherical homology groups in the persistent homology of four samples from the sphere, using each of three landmark procedures and three persistence computations. Similar plots of absolute dominance (not shown) tell a consistent story, but the distributions are more skewed so the comparisons are less clear.
\label{fig:sphere}
}
\end{figure}

Figure\nbs\ref{fig:sphere} compares the relative dominance of the
spherical homology groups in each case. When PH is computed using VR or
alpha complexes, maxmin better recovers the homology of the sphere
except on uniform samples, while lastfirst and random selection better
detect the void. Random selection is usually better than lastfirst
selection at detecting this void when samples are non-uniform, which
indicates that lastfirst selection still oversamples from less dense
regions. Lastfirst and maxmin perform similarly when PH is computed
using witness complexes.

\hypertarget{interpolative-nearest-neighbors-prediction-2}{%
\subsection{Interpolative nearest neighbors
prediction}\label{interpolative-nearest-neighbors-prediction-2}}

\begin{figure}
\includegraphics[width=\textwidth,trim=0 55 0 0,clip=true]{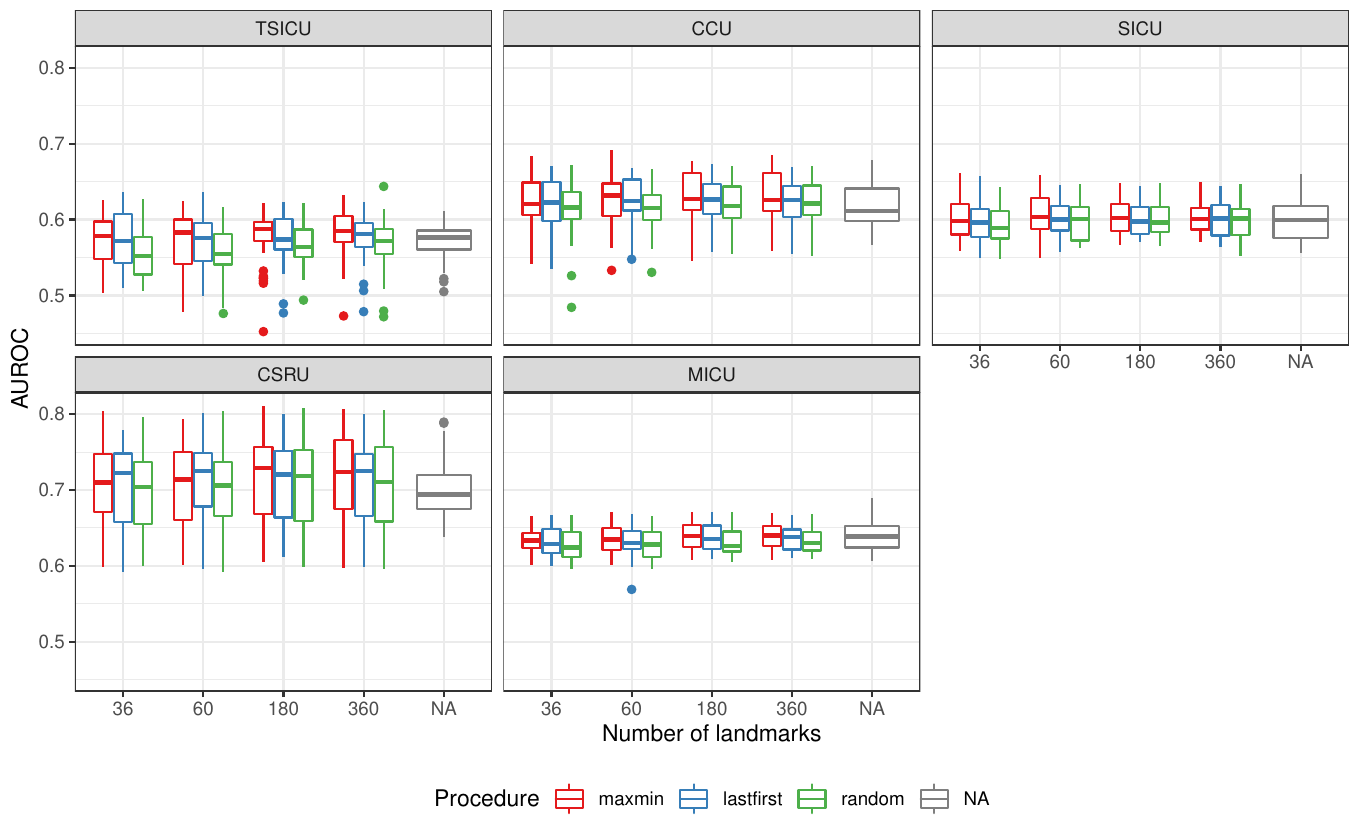}
\includegraphics[width=\textwidth]{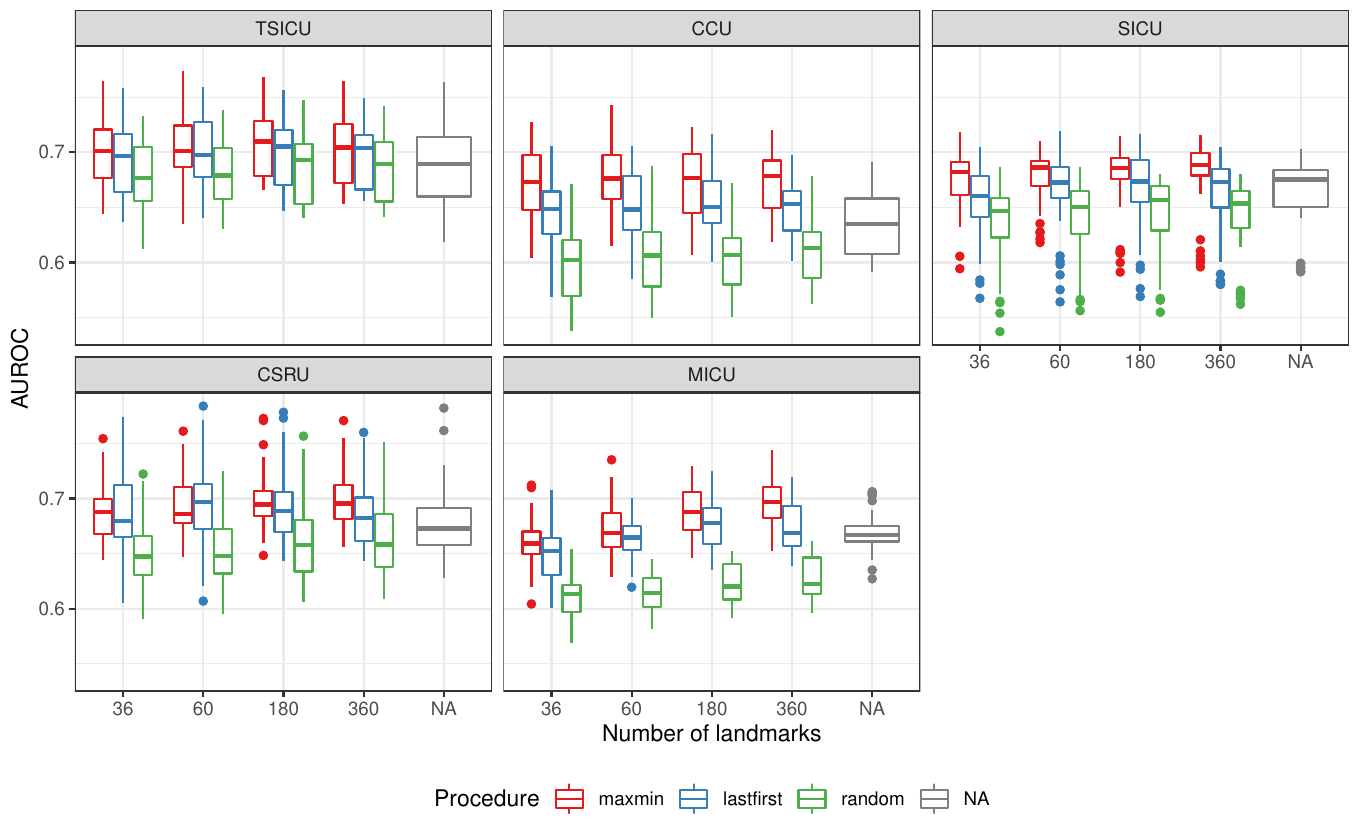}
\caption{
AUROCs of the interpolative predictive models of mortality in five MIMIC-III care units based on covers constructed using random, maxmin, and lastfirst procedures under angular distance on domain-informed variables (top) and on RT-transformed data (below) to generate landmarks.
Each boxplot summarizes AUROCs from $6 \times 6 = 36$ models, one for each combination of outer and inner fold.
AUROCs of simple nearest-neighbor predictive models are included for comparison.
\label{fig:knn-mimic-cos-rt}
}
\end{figure}

Maxmin-based ball covers and lastfirst-based neighborhood covers were
used to build interpolative nearest neighbors models to predict
mortality using data from MIMIC-III.
Figures\nbs\ref{fig:knn-mimic-cos-rt} compares their performance in each
critical care unit. Results differed greatly by distance measure and
less so by cover construction.

We used multiple linear regression quantified the relative importance of
care unit (population), distance measure, cover construction, and number
of landmarks. We modeled the expected AUROC for each folded
cross-validation step \(Y_i\) as a linear function of the care unit
\(U\) (reference value \(u_0 =\) TSICU), the distance measure \(D\)
(reference value \(d_0 =\) RT), the sampling procedure \(S\) (reference
value \(s_0 =\) random), and the number of landmarks \(L\) also treated
as a categorical variable (reference value \(\ell_0 = 36\)). We fit one
model with main effects only (Equation\nbs\ref{eqn:knn-mimic-main}) and
another model with two- and three-way interaction effects among unit,
measure, and procedure (Equation\nbs\ref{eqn:knn-mimic-prod}).

\begin{align}
\label{eqn:knn-mimic-main}
Y_i &= \beta_0 + \beta_U U_i + \beta_D D_i + \beta_S S_i + \beta_L L_i + \epsilon_i \\
\label{eqn:knn-mimic-prod}
Y_i &= \beta_0 + \beta_U U_i + \beta_D D_i + \beta_S S_i + \beta_U U_i \beta_D D_i + \beta_U U_i \beta_S S_i + \beta_D D_i \beta_S S_i + \beta_L L_i + \epsilon_i
\end{align}

\begin{figure}
\includegraphics[width=\textwidth]{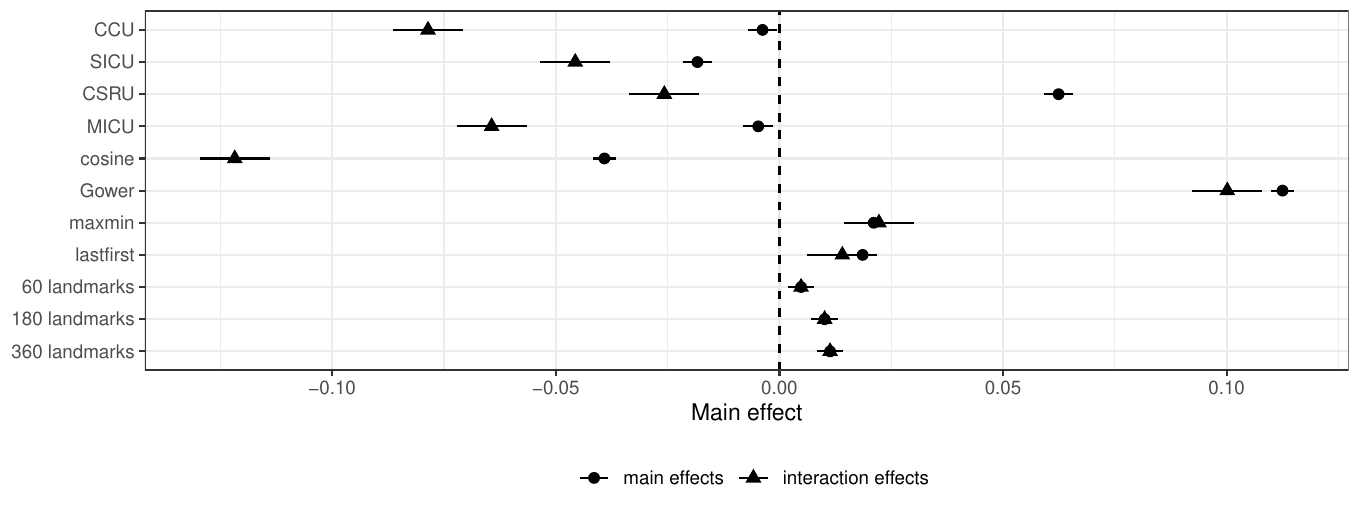}
\caption{
Main effects of the multiple regression models defined by Equations\nbs\ref{eqn:knn-mimic-main} and \ref{eqn:knn-mimic-prod}.
\label{fig:knn-mimic-main}
}
\end{figure}

Figure\nbs\ref{fig:knn-mimic-main} juxtaposes the main effects obtained
from both models. Recall that all models were optimized for neighborhood
size and weighting function before being evaluated. In both models,
overall performance was uncorrelated with unit size (by which the units
are ordered, with TSICU the smallest), Gower distance provided for
better-performing models than cosine distance, ball covers slightly
outperformed neighborhood covers, and performance improved with
additional landmarks.

\begin{figure}
\includegraphics[width=\textwidth]{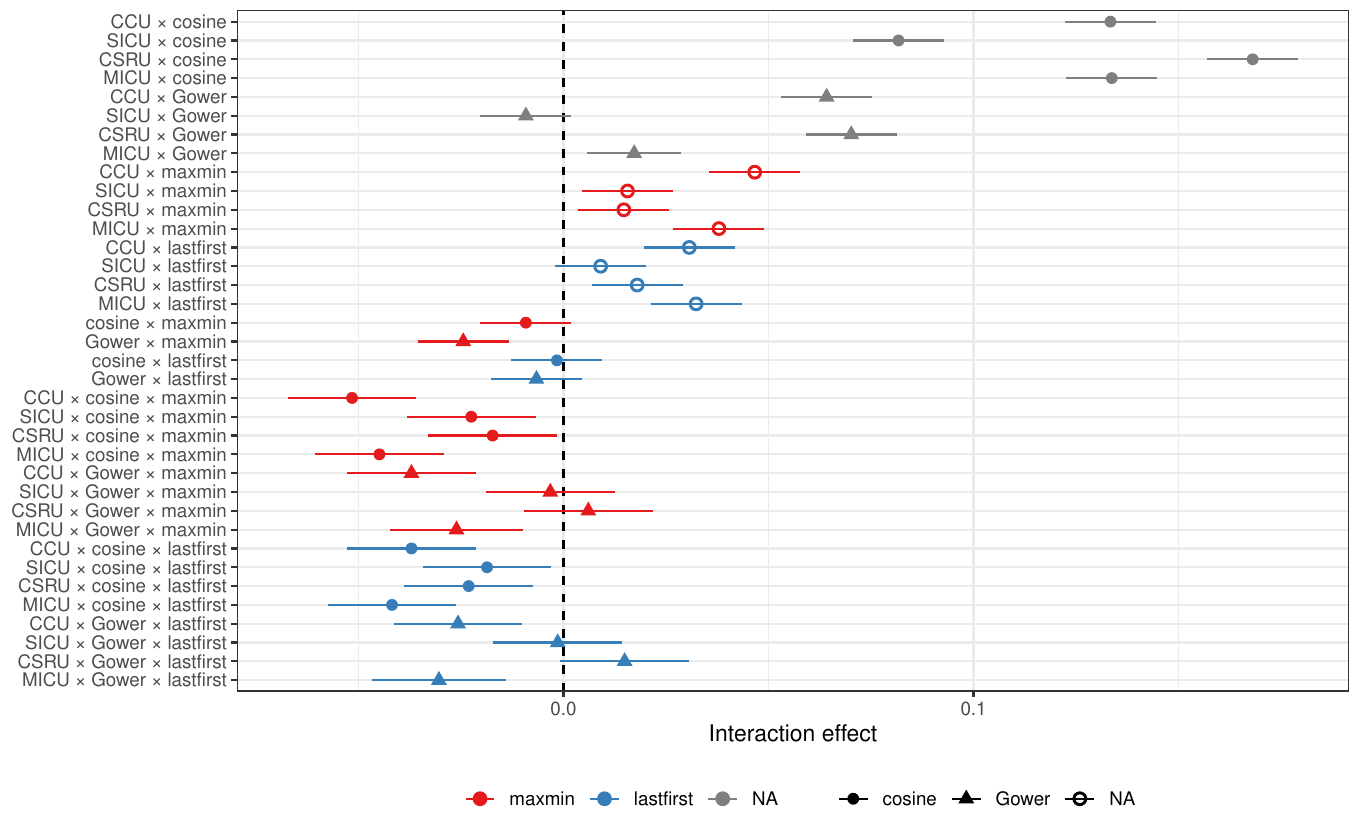}
\caption{
Interaction effects of the multiple regression model defined by Equations\nbs\ref{eqn:knn-mimic-prod}.
\label{fig:knn-mimic-prod}
}
\end{figure}

The lower main effect estimates of the models with interaction effects
compensate for the positive skew of the interaction effects themselves,
so we only draw comparisons within groups of interaction effects. The
Gower distance is clearly superior to cosine distance, and both
outperform the RT-transform cosine distance. Ball and neighborhood
covers perform similarly throughout and have similar dependencies on
care unit and distance measure.

\hypertarget{references}{%
\section*{References}\label{references}}
\addcontentsline{toc}{section}{References}

\hypertarget{refs}{}
\begin{CSLReferences}{1}{0}
\leavevmode\vadjust pre{\hypertarget{ref-Akkiraju1995}{}}%
Akkiraju, Nataraj, Herbert Edelsbrunner, Michael Facello, Ping Fu, Ernst
P. Mücke, and Carlos Varela. 1995. {``Alpha Shapes: Definition and
Software.''} \emph{The Geometry Center}.

\leavevmode\vadjust pre{\hypertarget{ref-Aziz2017}{}}%
Aziz, Rabia, C. K. Verma, Namita Srivastava, Rabia Aziz, C. K. Verma,
and Namita Srivastava. 2017. {``Dimension Reduction Methods for
Microarray Data: A Review.''} \emph{AIMS Bioengineering} 4 (1): 179--97.
\url{https://doi.org/10.3934/bioeng.2017.1.179}.

\leavevmode\vadjust pre{\hypertarget{ref-Becht2019}{}}%
Becht, Etienne, Leland McInnes, John Healy, Charles-Antoine Dutertre,
Immanuel W H Kwok, Lai Guan Ng, Florent Ginhoux, and Evan W Newell.
2019. {``Dimensionality Reduction for Visualizing Single-Cell Data Using
{UMAP}.''} \emph{Nature Biotechnology} 37 (1): 38--44.
\url{https://doi.org/10.1038/nbt.4314}.

\leavevmode\vadjust pre{\hypertarget{ref-Boissonnat2018}{}}%
Boissonnat, Jean-Daniel, Frédéric Chazal, and Mariette Yvinec. 2018.
\emph{Geometric and {Topological Inference}}. Cambridge {Texts} in
{Applied Mathematics}. {Cambridge}: {Cambridge University Press}.
\url{https://doi.org/10.1017/9781108297806}.

\leavevmode\vadjust pre{\hypertarget{ref-Bouguessa2006}{}}%
Bouguessa, Mohamed, Shengrui Wang, and Haojun Sun. 2006. {``An Objective
Approach to Cluster Validation.''} \emph{Pattern Recognition Letters} 27
(13): 1419--30. \url{https://doi.org/10.1016/j.patrec.2006.01.015}.

\leavevmode\vadjust pre{\hypertarget{ref-Bukatin2009}{}}%
Bukatin, Michael, Ralph Kopperman, Steve Matthews, and Homeira
Pajoohesh. 2009. {``Partial {Metric Spaces}.''} \emph{The American
Mathematical Monthly} 116 (8): 708--18.
\url{https://doi.org/10.4169/193009709X460831}.

\leavevmode\vadjust pre{\hypertarget{ref-ByczkowskaLipinska2017}{}}%
Byczkowska-Lipińska, Liliana, and Agnieszka Wosiak. 2017. {``Redukcja
Strumienia Danych Pozyskiwanych z Urządzeń Diagnostyki Medycznej Za
Pomocą Technik Selekcji Przypadków.''} \emph{Przegląd Elektrotechniczny}
93 (12): 115--18. \url{https://doi.org/10.15199/48.2017.12.29}.

\leavevmode\vadjust pre{\hypertarget{ref-Dai2020}{}}%
Dai, Leyu, He Zhu, and Dianbo Liu. 2020. {``Patient Similarity: Methods
and Applications.''} \emph{arXiv:2012.01976 {[}Cs{]}}, December.
\url{https://arxiv.org/abs/2012.01976}.

\leavevmode\vadjust pre{\hypertarget{ref-Dave1996}{}}%
Dave, Rajesh N. 1996. {``Validating Fuzzy Partitions Obtained Through
c-Shells Clustering.''} \emph{Pattern Recognition Letters} 17 (6):
613--23. \url{https://doi.org/10.1016/0167-8655(96)00026-8}.

\leavevmode\vadjust pre{\hypertarget{ref-deSilva2004}{}}%
de Silva, Vin, and Gunnar Carlsson. 2004. {``Topological Estimation
Using Witness Complexes.''} In \emph{{SPBG}'04 {Symposium} on
{Point-Based Graphics} 2004}, edited by Markus Gross, Hanspeter Pfister,
Marc Alexa, and Szymon Rusinkiewicz, 157--66. {The Eurographics
Association}.

\leavevmode\vadjust pre{\hypertarget{ref-Diaconis2013}{}}%
Diaconis, Persi, Susan Holmes, and Mehrdad Shahshahani. 2013.
{``Sampling from a {Manifold}.''} In \emph{Advances in {Modern
Statistical Theory} and {Applications}: {A Festschrift} in Honor of
{Morris L}. {Eaton}}, 10:102--25. Institute of {Mathematical Statistics
Collections}. {Institute of Mathematical Statistics}.

\leavevmode\vadjust pre{\hypertarget{ref-Dlotko2019}{}}%
Dłotko, Paweł. 2019. {``Ball Mapper: A Shape Summary for Topological
Data Analysis,''} January.

\leavevmode\vadjust pre{\hypertarget{ref-Eddelbuettel2011}{}}%
Eddelbuettel, Dirk, and Romain Francois. 2011. {``Rcpp: {Seamless R} and
{C}++ {Integration}.''} \emph{Journal of Statistical Software} 40 (8):
1--18. \url{https://doi.org/10.18637/jss.v040.i08}.

\leavevmode\vadjust pre{\hypertarget{ref-Falasconi2010}{}}%
Falasconi, M., A. Gutierrez, M. Pardo, G. Sberveglieri, and S. Marco.
2010. {``A Stability Based Validity Method for Fuzzy Clustering.''}
\emph{Pattern Recognition} 43 (4): 1292--1305.
\url{https://doi.org/10.1016/j.patcog.2009.10.001}.

\leavevmode\vadjust pre{\hypertarget{ref-Goldberger2000}{}}%
Goldberger, Ary L., Luis A. N. Amaral, Leon Glass, Jeffrey M. Hausdorff,
Plamen Ch. Ivanov, Roger G. Mark, Joseph E. Mietus, George B. Moody,
Chung-Kang Peng, and H. Eugene Stanley. 2000. {``{PhysioBank},
{PhysioToolkit}, and {PhysioNet}: {Components} of a {New Research
Resource} for {Complex Physiologic Signals}.''} \emph{Circulation} 101
(23). \url{https://doi.org/10.1161/01.CIR.101.23.e215}.

\leavevmode\vadjust pre{\hypertarget{ref-Gower1971}{}}%
Gower, J. C. 1971. {``A {General Coefficient} of {Similarity} and {Some}
of {Its Properties}.''} \emph{Biometrics} 27 (4): 857--71.
\url{https://doi.org/10.2307/2528823}.

\leavevmode\vadjust pre{\hypertarget{ref-Hester2020}{}}%
Hester, Jim. 2020. {``{bench}: {High Precision Timing} of {R
Expressions}.''}

\leavevmode\vadjust pre{\hypertarget{ref-Ivakhno2007}{}}%
Ivakhno, Sergii, and J. Douglas Armstrong. 2007. {``Non-Linear
Dimensionality Reduction of Signaling Networks.''} \emph{BMC Systems
Biology} 1 (1): 27. \url{https://doi.org/10.1186/1752-0509-1-27}.

\leavevmode\vadjust pre{\hypertarget{ref-Johnson2016}{}}%
Johnson, Alistair E. W., Tom J. Pollard, Lu Shen, Li-wei H. Lehman,
Mengling Feng, Mohammad Ghassemi, Benjamin Moody, Peter Szolovits, Leo
Anthony Celi, and Roger G. Mark. 2016. {``{MIMIC-III}, a Freely
Accessible Critical Care Database.''} \emph{Scientific Data} 3 (1).
\url{https://doi.org/10.1038/sdata.2016.35}.

\leavevmode\vadjust pre{\hypertarget{ref-Johnson2016b}{}}%
Johnson, Alistair, Tom Pollard, and Roger Mark. 2016. {``{MIMIC-III
Clinical Database}.''} {PhysioNet}.
\url{https://doi.org/10.13026/C2XW26}.

\leavevmode\vadjust pre{\hypertarget{ref-Johnston1976}{}}%
Johnston, J. W. 1976. {``Similarity Indices {I}: What Do They
Measure.''} {United States}.

\leavevmode\vadjust pre{\hypertarget{ref-Kachanovich2021}{}}%
Kachanovich, Siargey. 2021. {``Witness Complex.''} In \emph{{GUDHI User}
and {Reference Manual}}, 3.4.1 ed. {GUDHI Editorial Board}.

\leavevmode\vadjust pre{\hypertarget{ref-Konstorum2018}{}}%
Konstorum, Anna, Nathan Jekel, Emily Vidal, and Reinhard Laubenbacher.
2018. {``Comparative {Analysis} of {Linear} and {Nonlinear Dimension
Reduction Techniques} on {Mass Cytometry Data}.''} Preprint.
\url{https://doi.org/10.1101/273862}.

\leavevmode\vadjust pre{\hypertarget{ref-Lee2015}{}}%
Lee, Joon, David M. Maslove, and Joel A. Dubin. 2015. {``Personalized
{Mortality Prediction Driven} by {Electronic Medical Data} and a
{Patient Similarity Metric}.''} Edited by Frank Emmert-Streib.
\emph{PLOS ONE} 10 (5): e0127428.
\url{https://doi.org/10.1371/journal.pone.0127428}.

\leavevmode\vadjust pre{\hypertarget{ref-Major2020}{}}%
Major, Vincent J, Neil Jethani, and Yindalon Aphinyanaphongs. 2020.
{``Estimating Real-World Performance of a Predictive Model: A Case-Study
in Predicting Mortality.''} \emph{JAMIA Open} 3 (2): 243--51.
\url{https://doi.org/10.1093/jamiaopen/ooaa008}.

\leavevmode\vadjust pre{\hypertarget{ref-Maria2014}{}}%
Maria, Clément, Jean-Daniel Boissonnat, Marc Glisse, and Mariette
Yvinec. 2014. {``The {Gudhi Library}: {Simplicial Complexes} and
{Persistent Homology}.''} In \emph{Mathematical {Software} \textendash{}
{ICMS} 2014}, edited by Hoon Hong and Chee Yap, 167--74. Lecture {Notes}
in {Computer Science}. {Berlin, Heidelberg}: {Springer}.
\url{https://doi.org/10.1007/978-3-662-44199-2_28}.

\leavevmode\vadjust pre{\hypertarget{ref-Maria2021}{}}%
Maria, Clément, Paweł Dłotko, Vincent Rouvreau, and Marc Glisse. 2021.
{``Rips Complex.''} In \emph{{GUDHI User} and {Reference Manual}}, 3.4.1
ed. {GUDHI Editorial Board}.

\leavevmode\vadjust pre{\hypertarget{ref-Matthews1994}{}}%
Matthews, S. G. 1994. {``Partial {Metric Topology}.''} \emph{Annals of
the New York Academy of Sciences} 728 (1): 183--97.
\url{https://doi.org/10.1111/j.1749-6632.1994.tb44144.x}.

\leavevmode\vadjust pre{\hypertarget{ref-Park2006}{}}%
Park, Yoon-Joo, Byung-Chun Kim, and Se-Hak Chun. 2006. {``New Knowledge
Extraction Technique Using Probability for Case-Based Reasoning:
Application to Medical Diagnosis.''} \emph{Expert Systems} 23 (1):
2--20. \url{https://doi.org/10.1111/j.1468-0394.2006.00321.x}.

\leavevmode\vadjust pre{\hypertarget{ref-Reutlinger2012}{}}%
Reutlinger, Michael, and Gisbert Schneider. 2012. {``Nonlinear
Dimensionality Reduction and Mapping of Compound Libraries for Drug
Discovery.''} \emph{Journal of Molecular Graphics and Modelling} 34
(April): 108--17. \url{https://doi.org/10.1016/j.jmgm.2011.12.006}.

\leavevmode\vadjust pre{\hypertarget{ref-Rouvreau2021}{}}%
Rouvreau, Vincent. 2021. {``Alpha Complex.''} In \emph{{GUDHI User} and
{Reference Manual}}, 3.4.1 ed. {GUDHI Editorial Board}.

\leavevmode\vadjust pre{\hypertarget{ref-Singh2007}{}}%
Singh, Gurjeet, Facundo Mémoli, and Gunner Carlsson. 2007.
{``Topological {Methods} for the {Analysis} of {High Dimensional Data
Sets} and {3D Object Recognition}.''} In \emph{Eurographics {Symposium}
on {Point-Based Graphics}}, edited by M. Botsch, R. Pajarola, B. Chen,
and M. Zwicker. {The Eurographics Association}.
\url{https://doi.org/10.2312/SPBG/SPBG07/091-100}.

\leavevmode\vadjust pre{\hypertarget{ref-Skaf2022}{}}%
Skaf, Yara, and Reinhard Laubenbacher. 2022. {``Topological Data
Analysis in Biomedicine: {A} Review.''} \emph{Journal of Biomedical
Informatics} 130 (June): 104082.
\url{https://doi.org/10.1016/j.jbi.2022.104082}.

\leavevmode\vadjust pre{\hypertarget{ref-Viswanath2017}{}}%
Viswanath, Satish E., Pallavi Tiwari, George Lee, Anant Madabhushi, and
for the Alzheimer's Disease Neuroimaging Initiative. 2017.
{``Dimensionality Reduction-Based Fusion Approaches for Imaging and
Non-Imaging Biomedical Data: Concepts, Workflow, and Use-Cases.''}
\emph{BMC Medical Imaging} 17 (1): 2.
\url{https://doi.org/10.1186/s12880-016-0172-6}.

\leavevmode\vadjust pre{\hypertarget{ref-Wang2007}{}}%
Wang, Weina, and Yunjie Zhang. 2007. {``On Fuzzy Cluster Validity
Indices.''} \emph{Fuzzy Sets and Systems} 158 (19): 2095--2117.
\url{https://doi.org/10.1016/j.fss.2007.03.004}.

\leavevmode\vadjust pre{\hypertarget{ref-Yoon2020}{}}%
Yoon, Hee Rhang, and Robert Ghrist. 2020. {``Persistence by {Parts}:
{Multiscale Feature Detection} via {Distributed Persistent Homology}.''}
{arXiv}. \url{https://doi.org/10.48550/arXiv.2001.01623}.

\leavevmode\vadjust pre{\hypertarget{ref-Zhong2020}{}}%
Zhong, Haodi, Grigorios Loukides, and Robert Gwadera. 2020.
{``Clustering Datasets with Demographics and Diagnosis Codes.''}
\emph{Journal of Biomedical Informatics} 102 (February): 103360.
\url{https://doi.org/10.1016/j.jbi.2019.103360}.

\end{CSLReferences}

\bibliographystyle{unsrt}
\bibliography{lastfirst.bib}

\end{document}